# Coherent Phonons in Carbon Nanotubes and Graphene

J.-H. Kim[a,b], A.R.T. Nugraha[c], L.G. Booshehri[a], E.H. Hároz[a], K. Sato[c], G.D. Sanders[d], K.-J. Yee[b], Y.-S. Lim[e], C.J. Stanton[d], R. Saito[c], and J. Kono[a,f,*]

[a] *Department of Electrical and Computer Engineering, Rice University, Houston, Texas 77005, U.S.A.*
[b] *Department of Physics, Chungnam National University, Daejeon 305-764, Republic of Korea*
[c] *Department of Physics, Tohoku University, Sendai 980-8578, Japan*
[d] *Department of Physics, University of Florida, Gainesville, Florida 32611, U.S.A.*
[e] *Department of Nano Science & Mechanical Engineering and Nanotechnology Research Center, Konkuk University, Chungju, Chungbuk 380-701, Republic of Korea*
[f] *Department of Physics and Astronomy, Rice University, Houston, Texas 77005, U.S.A.*
[*] Corresponding author. *E-mail address:* kono@rice.edu

**Abstract.** We review recent studies of coherent phonons (CPs) corresponding to the radial breathing mode (RBM) and G-mode in single-wall carbon nanotubes (SWCNTs) and graphene. Because of the bandgap-diameter relationship, RBM-CPs cause bandgap oscillations in SWCNTs, modulating interband transitions at terahertz frequencies. Interband resonances enhance CP signals, allowing for chirality determination. Using pulse shaping, one can selectively excite specific-chirality SWCNTs within an ensemble. G-mode CPs exhibit temperature-dependent dephasing via interaction with RBM phonons. Our microscopic theory derives a driven oscillator equation with a density-dependent driving term, which correctly predicts CP trends within and between ($2n+m$) families. We also find that the diameter can initially increase or decrease. Finally, we theoretically study the radial breathing like mode in graphene nanoribbons. For excitation near the absorption edge, the driving term is much larger for zigzag nanoribbons. We also explain how the armchair nanoribbon width changes in response to laser excitation.

## 1. Introduction

The physical, chemical, and optical properties of crystalline solids are determined by the atomic-level interplay between light and the electrical and vibrational forces that tightly bind each atom inside the crystal lattice. Thus, a microscopic understanding of the dynamics and interactions of electrons, phonons, and photons is needed to correctly interpret macroscopic material properties and predict new phenomena. Recent progress in the fabrication of nanomaterials has been impressive and holds promise for future optoelectronic device applications. In order to assess their optimum capabilities, it is necessary to probe their microscopic properties under non-equilibrium conditions with ultrashort time resolution. With ultrafast laser spectroscopy, one can probe electronic and vibrational dynamics in real time. Numerous time-resolved detection techniques have been developed over the past few decades, and ultrafast phenomena can now be studied with a time resolution as short as 1 fs, which is shorter than one optical phonon period in most solids.

This article focuses on ultrafast optical phenomena in two carbon-based nanostructures with extraordinary properties: single-wall carbon nanotubes (SWCNTs) and graphene. With uniquely simple but unusual band structures, SWCNTs and graphene provide low-dimensional prototypes for studying the dynamics and interactions of electrons and phonons in one- and two-dimensions (1D and 2D), respectively. Recent continuous wave (CW) optical studies of SWCNTs and graphene have produced a world of intriguing phenomena, including strong Coulomb interactions enhanced by the low-dimensionality as well as interaction between excited electronic/excitonic states and phonons.

We have reported time-dependent observations of the lattice vibrations in these low-dimensional carbon structures [1-5]. Using femtosecond pump-probe spectroscopy, we observed coherent phonons (CPs) corresponding to the low-frequency RBM and the high-frequency G-mode. The observed phonon frequencies exactly correspond to those seen in traditional Raman spectroscopy in the same sample, but with narrower phonon linewidths, no photoluminescence signal or Rayleigh scattering background to obscure features, and high resolution allowing normally blended peaks to appear as distinct features. We found that CP signals are resonantly enhanced when the pump pulse resonantly excites excitons, allowing us to obtain precise information on chiralities present in a given SWCNT sample [4]. Furthermore, because the bandgap and diameter in SWCNTs are inversely proportional to each other, the bandgap coherently oscillates as the lattice undergoes coherent RBM oscillations [1]. This is a novel way of modulating interband optical absorption at terahertz (THz) frequencies and opens intriguing possibilities for novel THz devices. In addition, using tailored trains of femtosecond pulses, we selectively excited RBM CPs of specific-chirality SWCNTs within an ensemble sample [2]. For G-mode CPs in SWCNTs, we observed thermally activated dephasing with an activation energy that coincides with the RBM energy [6]. This suggests that the high-energy G-mode (i.e., optical phonons) can dephase via interaction with the RBM similar to the decay of the zone-center optic phonon into two zone edge acoustic phonons in GaAs. The dephasing time of the G-mode in graphene was found to be shorter than in semiconducting SWNCTs but longer than in metallic SWCNTs. In both SWCNTs and graphene, strong polarization dependence was observed in CP generation and detection [3, 5].

We have also developed a microscopic theory explaining our observations [7-10]. We find that the CP amplitudes satisfy a driven oscillator equation with a driving term depending on the photoexcited carrier density. We find that the RBM CP amplitude is a strong function of the photon excitation energy and polarization. In particular, we accurately predict the relative strengths of the CP signal for different chirality nanotubes [7, 9]. Furthermore, we predict that the nanotube diameter can initially either increase *or* decrease in response to femtosecond laser excitation. Finally, we developed a microscopic theory for CP generation and detection in armchair and zigzag graphene nanoribbons [10]. We examined the CP radial breathing like mode (RBLM) amplitudes as a function of excitation energy and nanoribbon type. For photoexcitation near the optical absorption edge, the CP driving term for the RBLM is much larger for zigzag nanoribbons where strong transitions between localized edge states provide the dominant contribution to the CP driving term. Using an effective mass theory, we explain how the armchair nanoribbon width changes in response to laser excitation.

This paper is organized as follows. In Section 2, we review the fundamental concepts of coherent phonon spectroscopy and give an overview of the basic optical and vibrational properties of carbon nanotubes and graphene. In Section 3, detailed descriptions are given of the experimental techniques used in our investigations. Section 4 presents experimental results, while Section 5 showcases some of the new predictions that have come out of our theoretical studies. Finally, we will summarize our findings and conclusions and provide an outlook for future studies and possible uses of coherent phonons.

## 2. Basic concepts

In this section, we provide an introduction to the theory of coherent phonon spectroscopy in semiconductor nanostructures including single-wall carbon nanotubes and graphene nanostructures. We also discuss the optical properties and Raman modes in graphitic materials.

*2.1 Coherent phonon spectroscopy*

Ultrafast femtosecond pump-probe spectroscopy is a useful tool for studying non-equilibrium carrier dynamics in a variety of semiconductor nanostructures since scattering times in these systems are in the 10-100s fs range. The decay of the reflection or transmission of the probe pulse as a function of delay time from the pump pulse provides information concerning details of the non-equilibrium carrier dynamics. Information obtained from these experiments includes: i) scattering rates, ii) electronic structure, and iii) many-body effects.

In addition to carrier dynamic effects, ultrafast pump-probe experiments have been shown to produce oscillating signals superimposed on the background carrier dynamics signal. These oscillations typically match one of the vibrational frequencies of the nanostructure and are known as *coherent phonons (CPs)* and the study of these oscillations is the field of CP spectroscopy [11-13]. Coherent optical phonons can be generated by the absorption of an ultrashort laser pulse whose duration is shorter than the period of the lattice vibration [14]. They are usually observed as periodic oscillations in the time-resolved differential reflectivity [15-20] or differential transmission. The oscillation frequency in transmission or reflection matches one of the phonon modes, which indicates that the phonon mode becomes coherently excited by the pulse.

Time-resolved CP spectroscopy is a powerful tool for investigating vibrational dynamics and allows direct measurement of the excited state phonon dynamics in the time domain, including phase information and dephasing times. To visualize the vibrational motion in a medium, light pulses with duration much shorter than the vibrational period are required.

*2.1.1 Coherent phonons*

When an ultrafast optical laser pulse rapidly creates electron-hole pairs across the bandgap in a semiconductor, the hot electrons and holes relax and lose energy primarily through the emission of optical and acoustic phonons. The phonons emitted, however, are *incoherent* phonons and not related to the oscillations observed in the differential and reflectivity spectra, although they can be responsible for the decay of the background signal [see Fig. 1(a)]. *Coherent* phonons are *quasi-classical* states formed from the coherent superposition of phonon harmonic oscillator eigenstates (i.e., the states with definite phonon number). Such a coherent phonon state is given by the superposition

$$\Psi^{coh} = |z\rangle = \sum_n \frac{z^n}{\sqrt{n!}} e^{-z^2/2} |n\rangle \qquad (2.1)$$

These states are essentially the same as those used in quantum optics to describe the quasi-classical quantized photon states of the electromagnetic field [21, 22]. The coherent phonon states are eigenfunctions of the phonon annihilation operator $b_q$ for phonons with wave vector $q$,

$$b_q |z\rangle = z |z\rangle, \qquad (2.2)$$

and represent minimum-uncertainty Gaussian wavepackets that oscillate back and forth in the parabolic potential without broadening, as shown in Fig. 1(b).

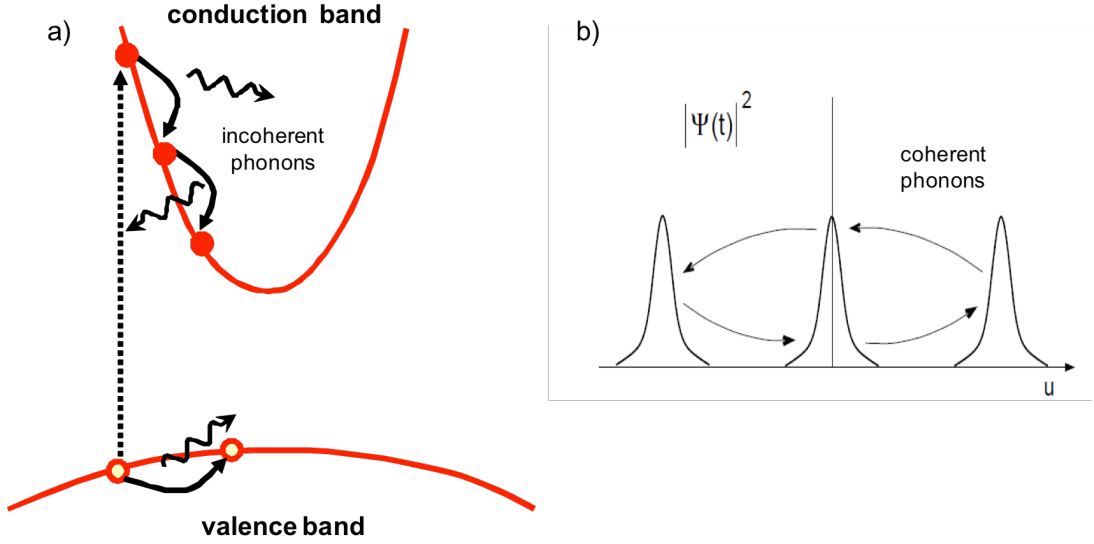

**Figure 1:** (a) Femtosecond laser excitation generates electron-hole pairs across the gap which relax and lose energy through phonon emission. These phonons are *incoherent* phonons and do not lead to oscillations in the differential reflectivity or transmission. (b) *Coherent* phonons are Gaussian wavepackets formed from the superposition of the eigenstates of the harmonic potential. The states oscillate back and forth in the harmonic potential without broadening.

*2.1.2 Coherent phonon generation mechanisms*

The excitation of the CP displacement amplitude $Q$ can be described phenomenologically as a harmonic oscillator driven by an external force, which depends on the electron and hole densities. When the femtosecond laser pulse rapidly creates electrons and holes across the gap, the force changes and triggers the coherent oscillations.

A phenomenological model for the CP amplitude $Q$, first introduced in Ref. [15], can be written as

$$\mu^*\left(\frac{\partial^2 Q(t)}{\partial t^2} + 2\gamma\frac{\partial Q(t)}{\partial t} + \Omega^2 Q(t)\right) = F^Q(n_e, n_h, t), \qquad (2.3)$$

where $\mu^*$ is a reduced lattice mass, $\gamma$ is a damping constant, $\Omega$ is the vibrational frequency, and $F^Q(n_e, n_h, t)$ is a driving force that depends on time through the electron and hole densities $n_e$ and $n_h$ that change rapidly with the laser pump pulse. The second term in Eq. (2.3) is a damping term, and the damping constant $\gamma$ is related to the dephasing time $T_2$ of the coherent mode via $\gamma = 1/T_2$. The notion of a dephasing time has been established for coherent excitations; in a density matrix representation, $T_2$ describes the temporal evolution of the non-diagonal terms of the density matrix. This dephasing time $T_2$ is related to the population decay time $T_1$, which describes the decay of the diagonal terms of the density matrix, via $2/T_2 = 1/T_1 + 1/T_p$, where $T_p$ is the decay time for truly phase-destroying processes [23] and can usually be neglected for vibrational excitations in solids. Therefore, the observed dephasing time of a coherent phonon mainly reflects the population decay time; that is, the observed dephasing times in a time-resolved experiment should give the same values as derived from non-time-resolved inelastic light-scattering experiments.

Equation (2.3) can be solved exactly in two limiting cases: a) an *impulsive* force

$$F(t) = I\delta(t) \tag{2.4}$$

where $I = \int F(t)\, dt$ is the total impulse delivered to the oscillator, and b) a *displacive* force

$$F(t) = \frac{F}{\mu^*}\theta(t) \tag{2.5}$$

where $\theta(t)$ is the Heavyside step function.

The solution for an impulsive force is given by:

$$Q(t) = \frac{I}{\mu^*\sqrt{\Omega^2 - \gamma^2}} e^{-\gamma t} \sin\left(\sqrt{\Omega^2 - \gamma^2}\, t\right). \tag{2.6}$$

An impulsive force would be similar to ringing a bell, where the driving force is applied for only a short time. An impulsive force results if the femtosecond pump pulse is not resonant with the conduction and valence band states as is the case in below bandgap excitation. Because of the uncertainty principle, *virtual* carriers are created and the density of these carriers will adiababitcally follow the pulse envelope.

The solution for a discplacive force is given by:

$$Q(t) = \frac{F}{\mu^*\Omega^2}\left[1 - \left(\frac{\gamma}{\sqrt{\Omega^2 - \gamma^2}}\sin\left(\sqrt{\Omega^2 - \gamma^2}\, t\right) + \cos\left(\sqrt{\Omega^2 - \gamma^2}\, t\right)\right)e^{-\gamma t}\right]. \tag{2.7}$$

A displacive force is analogous to putting weights on a spring suspended from the ceiling. The weights cause the spring to stretch to a new equilibrium position and if the weights were applied fast enough, the spring will oscillate around the new equilibrium position. Displacive forces typically arise when the femtosecond pump pulse has resonant transitions creating *real* carriers in the semiconductor as is the case for excitation above the bandap. In this case, the driving force would be proportional to the integral in time of the pulse envelope, which is approximately given by a step function.

Results for impulsive and displacive excitation are shown in Figs. 2(a) and 2(b), respectively. Note that the displacive transition moves to a new equilibrium position and that there is about a ¼ period phase lag. Due to the broad spectral width for ultrafast pump excitation, there are Fourier components for exitation both above and below the bandgap and both real and virtual carriers will in general be created leading to impulsive and displacive contributions to the generation process. The impulsive contribution, as discussed below in the section on impulsive transiently stimulated Raman processes will usually have a strong angular dependence on the polarization of the pump pulse.

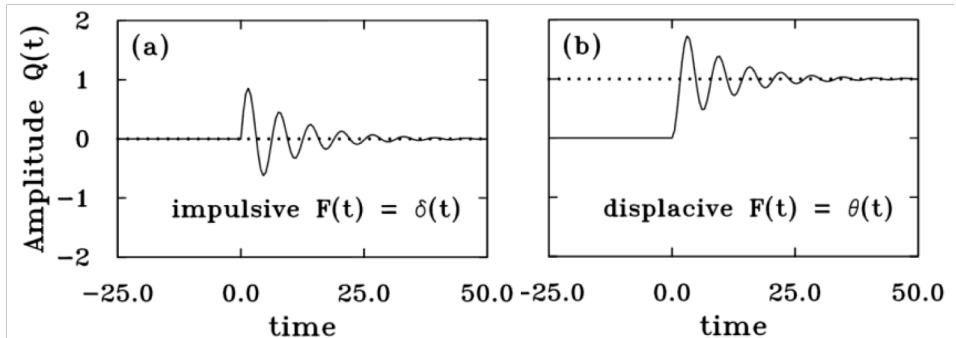

**Figure 2:** Solution to Eq. (2.3) for (a) an impulsive and (b) displacive driving force. The damping is given by $\gamma = 0.1\Omega$.

*2.1.3 Microscopic origin of the oscillator model*

The key to understanding the microscopic origin of the driven oscillator model for coherent phonon generation is to realize that the coherent phonon amplitude is proportional to the expectation value

$$Q(t) \sim \langle b_q + b^\dagger_{-q} \rangle, \qquad (2.8)$$

where $\langle b^\dagger_{-q} \rangle$ and $\langle b_q \rangle$ are the expectation values of the phonon creation and annihilation operators for phonon wavector $-q$ and $q$, respectively. Note that the coherent amplitude is proportional to the Fourier components of the displacement and that the expectation values do not vanish for the coherent states given by Eq. (2.1). In a state with a definite number of phonons (i.e., eigenstates of the harmonic oscillator), however, the expectation value is identically zero. In a coherent state, given by Eq. (2.1), the coherent amplitude can be non-zero.

To obtain the equations of motion for $\langle b^\dagger_{-q} \rangle$ and $\langle b_q \rangle$, one uses the Heisenburg equations for the expectation values of operators by commuting with the Hamilitonian:

$$\hat{H} = \underbrace{\sum_{\mathbf{k}} \varepsilon_{\mathbf{k}} c^\dagger_{\mathbf{k}} c_{\mathbf{k}}}_{\text{electron}} + \underbrace{\sum_q \hbar \Omega_q b^\dagger_q b_q}_{\text{phonon}} + \underbrace{\sum_{\mathbf{k},q} M_{\mathbf{k},q} \left( b^\dagger_{-q} + b_q \right) c^\dagger_{\mathbf{k}} c_{\mathbf{k}+q}}_{\text{electron-phonon}} \qquad (2.9)$$

where the first term is the electron part of the Hamiltonian, the second term is the phonon Hamiltonian, and the third term is the electron-phonon interaction Hamiltonian where $M_{k,q}$ is the electron-phonon matrix element for electron wave vector **k** and phonon wavevector $q$.

After some algebra, one obtains the equation of motion for the coherent amplitude given in Eq. (2.8):

$$\frac{\partial^2}{\partial t^2} Q_q + \Omega_q^2 Q_q = -2\Omega_q \sum_{\mathbf{k}} M_{\mathbf{k}q} n_{\mathbf{k},\mathbf{k}+q}. \qquad (2.10)$$

There is no damping term here since the anharmonic terms in the electron-phonon Hamiltonian are neglected. Note that this closely resembles the phenomenological model given in Eq. (2.3) and provides an expression for the driving term. Here, $n_{\mathbf{k},\mathbf{k}+q}$ is the Fourier transform of the electronic density matrix

$$\rho_{\text{el}}(\mathbf{r}) = \frac{e}{V} \sum_{\mathbf{k},q} n_{\mathbf{k},\mathbf{k}+q} \, e^{iqr}. \qquad (2.11)$$

The rapid creation of electrons/holes by the femtosecond pump pulse changes the forcing function and triggers the coherent phonon oscillations.

Since the wavelength of the pump laser is large compared with the spacing between atoms in the nanostructure, *usually* the electrons and holes are created in a macroscopicaly uniform state, which excites only the $q \approx 0$ phonon modes. However, we note that there are some cases where one can excite $q \neq 0$ modes. This can occur in: a) superlattices, where the light is absorbed only in the wells leading to coherent acoustic modes at the *superlattice* wavevector, b) epilayers, where absorption may occur only in a few layers near the surface and can lead to propagating acoustic phonon wavepackets, which can modulate the transmission and reflection of the probe pulse, and c) with multiple pump pulses that can create transient gratings.

To date, coherent phonon studies in carbon nanotubes and graphene nanostructures have focused primarily on the $q = 0$ phonon modes. These are the RBM and G-modes in carbon nanotubes, the iTO and LO (G) modes in graphene, and the RBLM modes in graphene nanoribbons. In addition, D-band (defect oriented phonon mode) corresponding to a $q \neq 0$ phonon mode coupled with elastic scattering of electrons with $-q$ wavevectors have also been observed by CP spectrscopy in SWCNTs [24] and graphene [25]. One expects that, in future studies, novel ways to excite *other* coherent $q \neq 0$ modes in these systems will be developed, yielding interesting results analogous to the case of propagating coherent acoustic phonon modes observed in traditional semiconductors. A diagram showing the vibrational directions for (a) the radial breathing mode and (b) the tangential G-modes in a SWCNT is shown in Fig. 8.

For a carbon nanotube, the notation is slightly more cumbersome. The equations of motion for CP modes are obtained using a microscopic description of the electron-phonon interaction. For each phonon mode in the nanotube, the CP amplitude is given as [26]

$$Q_{\beta v q}(t) = \langle b_{\beta v q}(t) + b^{\dagger}_{\beta v, -q}(t) \rangle \tag{2.12}$$

where the subscript $\beta$ labels the six phonon modes in the graphene phonon-dispersion relation, $v$ is the cutting line index (1D Brillouin zones shown in the 2D Brillouin zone of graphene), and $q$ is the phonon wavevector. Equations of motion for $Q_{\beta v q}(t)$ can be obtained from the phonon and electron-phonon Hamiltonians, and the only excited CP modes are the $v = q = 0$ modes whose amplitudes satisfy a driven oscillator equation,

$$\frac{\partial^2 Q_{\beta v q}}{\partial t^2} + \Omega^2_{\beta v q} Q_{\beta v q}(t) = S_{\beta v q}(t) \tag{2.13}$$

which closely resembles the phenomenological Eq. (2.3). Again, there is no damping term here due to neglect of anharmonic terms in the electron-phonon Hamiltonian. In what follows we drop the $v = q = 0$ subscripts.

The CP driving function $S_\beta(t)$ depends on the photoexcited electron distribution functions. Sanders *et al.* solved for the time-dependent distribution functions in the Boltzmann equation formalism, taking photogeneration and relaxation effects into account [7]. In CP spectroscopy, an ultrafast laser pulse generates carriers on a time scale short in comparison with the CP period. An ultrashort pulse excites coherent vibrational waves through an Impulsive Stimulated Raman Scattering (ISRS) process treated in the following section. In our experiments, we use about 12 fs (60 fs) ultrafast laser pulses to excite G-mode (RBM) CPs with oscillation periods of around 21 fs (140 fs).

After photoexcitation, the carriers scatter and recombine. The driving function $S_\beta(t)$ rises sharply on the time scale of the pump pulse envelope (~ 10's of fs) and then vanishes more slowly on the carrier relaxation time scale (~several ps). The rapid initial jump in $S_\beta(t)$ gives rise to an oscillatory part of the coherent phonon amplitude $Q_\beta(t)$ at the CP frequency $\Omega_\beta$. The observed CP signal is proportional to the power spectrum of the oscillatory part of $Q_\beta(t)$. The photogeneration rate in the Boltzmann equation depends on the polarization of the incident ultrafast laser pulse. Optical absorption due to light polarized parallel to the tube axis is greater than absorption due to light polarized perpendicular to the axis because of the depolarization effect for perpendicular polarization [27, 28].

*2.1.4 Detection of coherent phonons*

The detection mechanism for CP oscillations in differential reflectivity or transmission can vary dramatically, depending on the system one is studying. For instance, in polar materials such as GaAs, both the electrons/holes and phonons can contribute to the observed oscillations. In this case, we solve the coupled plasmon-phonon mode equations and both modes can contribute to the oscillatory response. At high densities, the phonon modes are screened by the photoexcited carriers, and the frequency of oscillations shifts. When the photoexcited carriers are not generated uniformly in the sample but are only generated typically with a Gaussian profile within the pump laser spot and within an absorption depth of the surface, we have to consider carrier diffusion effects [29-32], as well as reducing the effect of the plasma contribution to the signal since it is density dependent and varies with position in the spot [14].

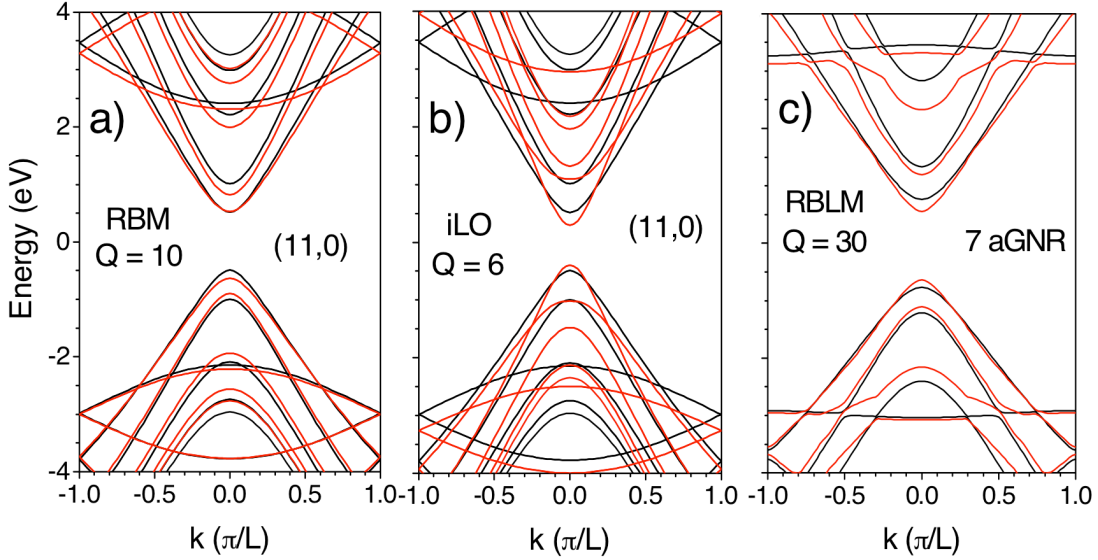

**Figure 3:** (a) The electronic structure in a (11,0) SWCNT. The black lines are the unstrained bands. The red lines represent the effect of the RBM expanding deformation. (b) The elecrtronic structure in a (11,0) SWCNT showing the effect of an LO phonon deformation (ref lines). (c) Similar plots of the electronic structure for a 7 aGNR. The red lines are for a RBLM distortion. The magnitude of the distortions are exaggerated.

In the case SWCNTs, the coherent RBM oscillation changes the tube diameter as a function of time. Since the energy bandgap of seminconducting SWCNTs is inversely proportional to the tube diameter, the CP RBM mode effectively modulates the bandgap on an ultrafast time scale. For incoherent RBM phonons whose relative phase is random, the energy bandgap does not vary with time. The effect of coherent RBM diameter oscillations on the SWCNT bandstructure is illustrated in Fig. 3, where we show the electronic structure for a (11,0) SWCNT [Fig. 3(a)] for two lattice deformations. The black lines represent the unstrained electronic structure and the red lines represent the deformed bands due to an expanding coherent RBM mode at fixed value of the coherent phonon amplitude, $Q$. Figure 3(b) shows the electronic structure for (11,0) SWCNT strained by an LO mode (red lines) while Fig. 3(c) shows similar curves for a 7 armchair graphene nanoribbon.

As can be seen from Fig. 3, near the band edge, the main effect of the RBM (or RBLM for the aGNR) is to approximately change the bandgap. The effect on the coefficient of absorption $\alpha$ for small changes in the gap is therefore:

$$\alpha(E-E_g) \approx \alpha(E-E_g^0) - \frac{\partial \alpha(E-E_g^0)}{\partial E} \delta E_g. \qquad (2.14)$$

Since the gap oscillates with the CP frequency, $\delta E_g \sim \delta E_g^{max} \cos(\Omega t + \phi)$, we see that the change in the absorption of the probe pulse due to the CP generated by the pump pulse is

$$\Delta \alpha \approx -\frac{\partial \alpha(E-E_g^0)}{\partial E} \delta E_g^{max} \cos(\Omega t + \phi). \qquad (2.15)$$

We thus see that for the RBM mode in SWCNT and the RBLM modes in graphene nanoribbons, CP spectroscopy is a *derivative spectroscopy* where $\alpha' \equiv \partial \alpha / \partial \varepsilon$ is observed. The actual signal measured in the SWCNT CP experiments is the power spectrum in which the change of the intensity for transmitted light is observed. Hence, the CP signal should be proportional to

$$CP \sim \left| \frac{\partial \alpha(E - E_g^0)}{\partial E} \right|^2. \quad (2.16)$$

In Fig. 4, we plot the absorption spectra for a 0-D system (single, *discrete* two level system like an exciton), a 1-D system $\left(\alpha \propto 1/\sqrt{E}\right)$, a 2-D system ($\alpha \sim $ constant) and a 3-D system $\left(\alpha \propto \sqrt{E}\right)$ in the top row. The curves are convoluted with a Lorentzian to take into account the linewidths of the transitions. We next show the negative of the derivative of the absorption with respect to energy $-\partial \alpha / \partial E$ in the middle row. Finally, we show the absolute square $|\partial \alpha / \partial E|^2$ which is approximately what would be measured in a CP experiment for the RBM and RBLM modes in the bottom row. We note the double peaked structure for both the 0-D and 1-D systems because of the singularity of α.

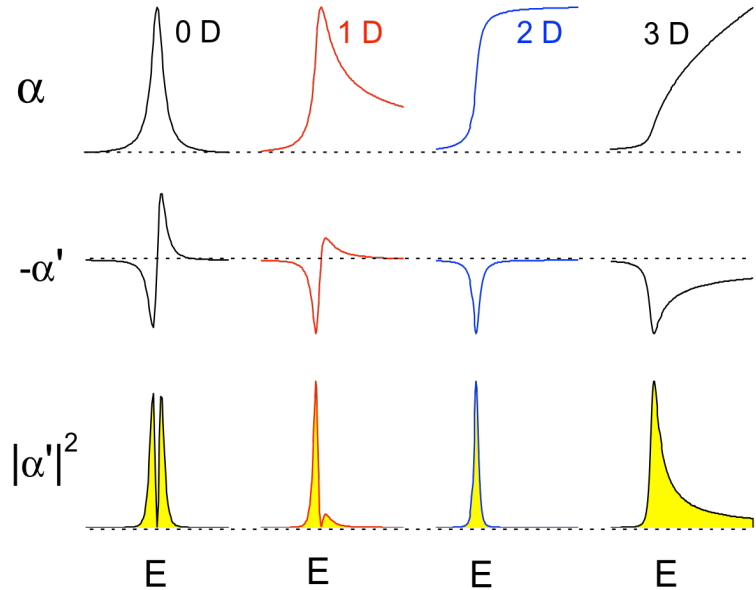

**Figure 4:** The absorption spectra $\alpha$, derivative of the absorption spectra $-\alpha'$ and $|\alpha'|^2$ for a 0D, 1D, 2D, and 3D system. The absorption curves have been convoluted to take into account lifetime broadening of the states.

*2.2 Phonons in graphene and carbon nanotubes*

Here we briefly describe the graphene phonon dispersion relations. Since the unit cell of monolayer graphene contains two carbon atoms, there are six phonon dispersion relations of which three are acoustic branches (A) and three are optic phonon branches (O). Further, two of the six phonon branches are longitudinal (L) modes and the remaining four are tangential (T) modes. In a L (T) phonon mode, the vibration direction is parallel (perpendicular) to the wave propagating direction as specified by the phonon wavevector $q$. The four T phonon modes consist of two in-plane phonon modes (iTA and iTO), whose vibration direction is within the graphene plane, and two out-of-plane phonon modes (oTA and oTO), whose vibration direction is out of the graphene plane. Since LA and LO phonon modes are always in-plane phonon modes, we do not need "i" for LA and LO.

Therefore, along the high symmetry ΓM and ΓK directions, the six phonon dispersions are the LO, iTO, oTO, LA, iTA, and oTA phonon modes, as shown in Fig. 5. In solids, only $q = 0$ phonon modes (zone center phonon modes) are observed as a first-order Raman process. According to group theory [33], the degenerate zone-center LO and iTO phonon modes belong to the two-dimensional $E_{2g}$ representation, and they are Raman active modes known as the G band [33], [34]. The oTO phonon modes are not Raman active (but infra-red active) phonon mode since the area of the unit cell does not change by the vibration and thus there is no electron-phonon matrix element. If we consider second-order Raman processes, the restriction of $q = 0$ is relaxed, and non-zone-center phonon modes can become Raman active. For example, the phonon modes around the K-point are important, since the D-band and G'-band are related to phonon modes in the vicinity of the K-point. Usually, second-order Raman process may give a broad and weak Raman signal. However, if we have a so-called double resonance process in which two intermediate states are resonant (energy and momentum conserved) states, then the double resonance Raman signal is sharp and strong compared with the first-order Raman process [35].

In the presence of free carriers (electrons and holes) at the Fermi energy, the frequency of the LO and iTO phonon branch near the Γ point and iTO phonon branch near the K-point become soft and broad because of the virtual excitation of electron-hole pairs due to the electron-phonon interaction with free carriers (the Kohn anomaly) [36]. When one applies a gate voltage to graphene, the phonon frequency can be changed as a function of gate voltage (or the Fermi energy) [37, 38].

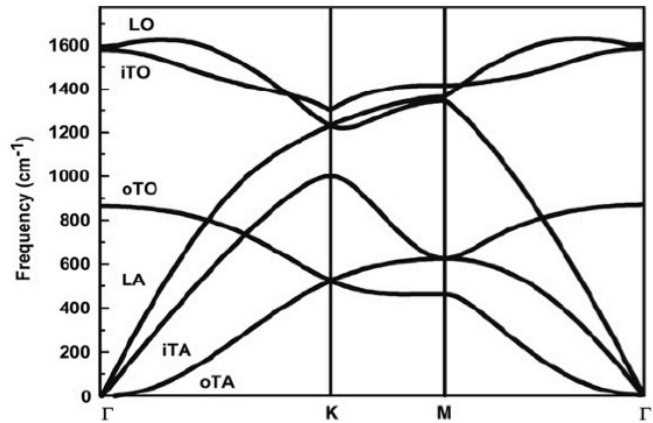

**Figure 5:** Calculated phonon dispersion relations of graphene showing the LO, iTO, LA, iTA and oTA phonon branches. Reproduced from Ref. [39].

Two notable Raman modes of monolayer graphene are the G-band appearing at around 1582 cm$^{-1}$ and the G' (or 2D) band at about 2700 cm$^{-1}$ [40]. In Fig. 6, using laser excitation at 2.33 eV, the Raman spectrum shows four prominent peaks for (a) single- and (b) double-layer graphene. The G-band peak is a first-order intra-valley Raman scattering, which corresponds to the $E_{2g}$ phonon (LO and iTO modes) at the Γ point [Fig. 7(b)]; the G band can be seen in any graphene related materials. The D band around 1350 cm$^{-1}$ are second-order intervalley scattering processes (scattering from K to K' (or K' to K) Dirac cones in the 2D Brillouin zone), which consists of iTO phonon emission and an elastic scattering due to the lattice defect with $q$ vectors connecting K and K' points [Fig. 7(a)]. Because of the double resonance condition for $q$, the D band frequency increases by 54 cm$^{-1}$ with increasing laser energy by 1 eV, which is known as the dispersion effect. The G′ band is a two-phonon, inter-valley, double resonance process of the iTO phonon mode near the K point [Fig. 7(c)], whose dispersion is 108 cm$^{-1}$/eV. It is important to note that the G' band is *not* an overtone mode of the D band since defect scattering is (not) involved for the latter (former). In fact, the G' band appears in a high-quality crystal without any defects, and the G' band frequency is slightly smaller than the twice of the D band frequency for a given laser energy. Finally, the overtone of the G band is located 3248 cm$^{-1}$, which is more than twice the energy of the G band. Scattering mechanisms of each Raman mode D, G, and G' are shown in Fig. 7.

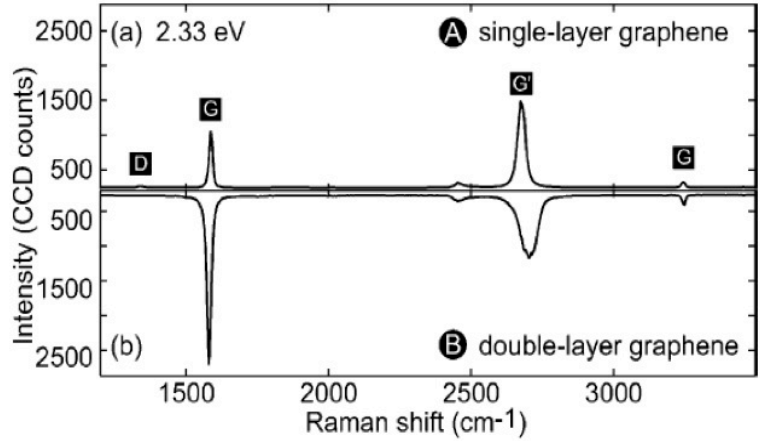

**Figure 6:** Raman spectrum of (a) single- and (b) double-layer graphene with excitation energy of 2.33 eV. Reproduced from Ref. [40].

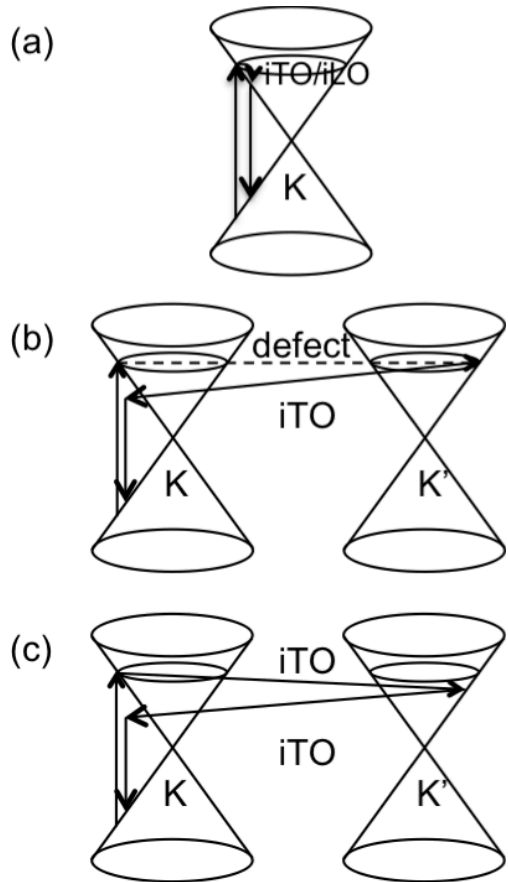

**Figure 7:** Scattering mechanisms of photo-excited carriers for the various Raman modes in graphene: (a) First-order G-band process. (b) One-phonon second-order double resonance process

for D-band. (c) Two-phonon second-order double resonance process for G' band. The dashed line shows defect-related elastic scattering, whereas solid lines show inelastic scattering.

In the case of SWCNTs, the phonon wave vector $q$ becomes discrete via the periodic boundary condition around the circumferential direction while in the direction of the nanotube axis $q$ is continuous as in one-dimensional materials. Thus, a zone folding procedure, similar to that for electronic band dispersions, is applied to obtain the phonon dispersion relations and phonon density of states for SWCNTs from those of the 2D graphene sheet [41]. Contrary to 2D graphene, the SWCNT shows many phonon branches in the phonon dispersion curve and a number of one-dimensional singularities in the phonon density of states appear (1D van Hove singularities). Raman modes of a SWCNT are similar to those of graphene, but there is a unique Raman mode in SWCNT, called the radial breathing mode (RBM), in which the diameter of a SWCNT, $d_t$, vibrates as shown in Fig. 8(a). The RBM phonon frequency for (bundled) SWCNTs is generally expressed by a simple formula,

$$\omega_{RBM} = A/d_t + B \qquad (2.15)$$

where the fitting parameters $A$ and $B$ are given by 223.5 nm·cm$^{-1}$ and 12.5 cm$^{-1}$, respectively [42], corresponding to Raman shifts of 100 cm$^{-1}$ < $\omega_{RBM}$ < 500 cm$^{-1}$ for SWCNTs with diameters of 0.4 nm < $d_t$ < 3 nm.

The assignment of ($n,m$) of a SWCNT is given by resonant Raman spectroscopy [43], in which one compares the laser energy and Raman shift of RBM with the calculated results of resonant $E_{ii}$ and phonon frequency, respectively. If a laser energy is resonant with $E_{ii}$, then the Raman signal of a particular ($n,m$) SWCNT species becomes strong (resonance Raman effect) even in a chirality-mixed sample. Combined with the information of $\omega_{RBM}$ from either Raman spectroscopy [44] or coherent phonon spectroscopy [45], one can know the value of $d_t$ and hence identify the index $i$ in $E_{ii}$ from the empirical Kataura plot ($E_{ii}$ values of all ($n,m$) SWCNTs as a function of $d_t$) [8, 46].

The G-mode in a SWCNT, which is the $E_{2g}$ mode of 2D graphene, is split into two – the G$^+$ (higher frequency) and G$^-$ (lower frequency) modes – because of the curvature of the cylindrical structure. In semiconducting SWCNTs, G$^+$ (G$^-$) corresponds to the LO (iTO) phonon mode, in which the vibration direction is parallel (perpendicular) to the nanotube axis [see Fig. 8 (b)]; the separation between the G$^+$ and G$^-$ modes is inversely proportional to the square of $d_t$ [47]. The relative intensities of the G$^+$ and G$^-$ peaks depend on the chiral angle of the SWCNT [48]. In addition to the two main peaks of G$^+$ and G$^-$, which have $A$ symmetry of the $C_n$ point group for a chiral SWCNT [33], $E_1$ and $E_2$ symmetry of the $C_n$ point group are also Raman active, although their peak intensities are expected to be much weaker.

In the case of a metallic SWCNT, the electron-phonon interaction acts only on the LO phonon mode, and the G$^-$ peak is now the LO phonon mode. The G$^-$ peak becomes broad and soft because of the Kohn anomaly effect [47, 49]. Thus, by changing the gate voltage, the G$^-$ peak changes drastically [38] and chirality dependent [50]. Phonon spectra and the intensities of resonance Raman and CP spectral features can be changed by many techniques by laser excitation energy, stress, magnetic field, polarization direction of light, gate voltage, temperature, and the dielectric constant of surrounding materials [51].

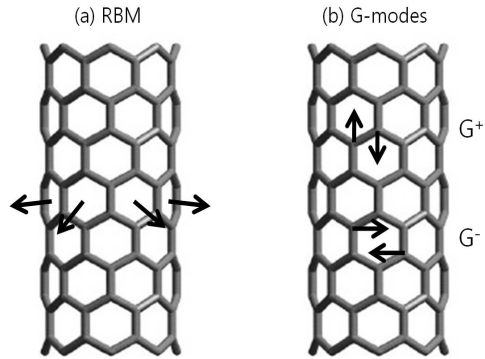

**Figure 8:** Schematic pictures of vibrational directions for (a) the radial breathing mode and (b) the tangential G-modes.

## 3. Experimental methods

Figure 9 schematically shows the degenerate pump-probe spectroscopy experimental setup that we used for coherent phonon studies of SWCNTs and graphene. The source was a Ti:sapphire laser oscillator with a pulse duration of 12 fs, which is shorter than one period of the G-mode oscillations in SWCNTs (21 fs). The output beam from the laser was divided into pump and probe beams by a 7:3 beam splitter, and the two laser beams overlapped at the same spot on the sample surface. The probe beam was delayed in time with respect to the pump beam by changing the optical path length using a computer-controlled delay stage. All experiments were done in transmission geometry, where the differential transmission of the probe beam was measured as a function time delay.

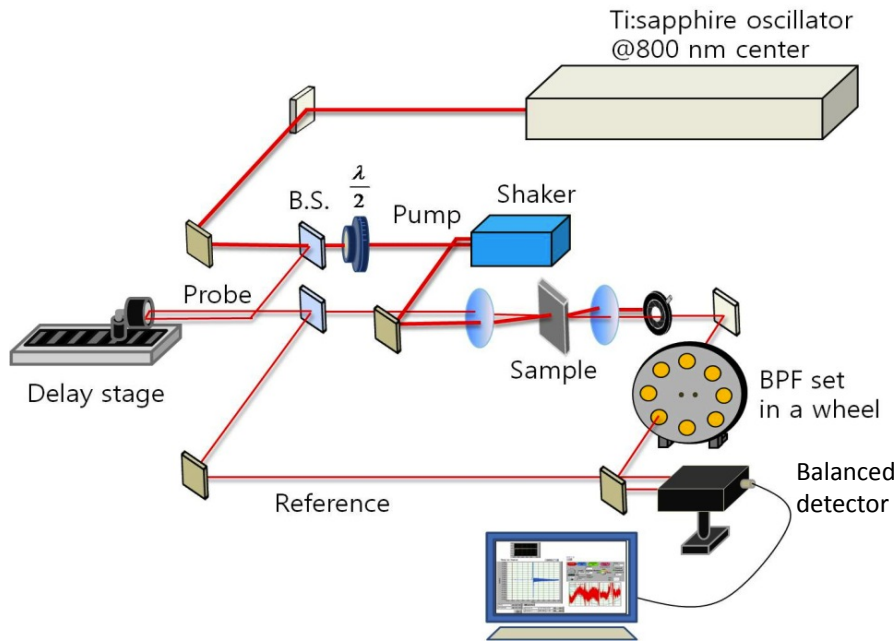

**Figure 9:** Degenerate pump-probe spectroscopy experimental setup. The half-wave plate (λ/2) in the pump beam path is used for measuring pump-polarization dependence, and the band-pass-filter (BPF) set in a wheel is used for spectrally resolved detection. The difference between the probe and reference beams is measured by the balanced detector.

The probe beam was further split in two by a 9:1 beam splitter, and the weaker beam used as the reference beam for the balanced detector. A shaker was used as a fast-scanning delay line for producing real-time signals. We could capture a time window of 5 ps with the shaker amplitude of 0.7 mm and ten scans per second by moving the shaker at 10 Hz. The half-wave plate in the pump beam path was used for rotating the pump-polarization with respect to the probe polarization in polarization-dependent measurements. The band-pass-filter (BPF) set, housed in a motorized wheel in front of the detector, was used for spectrally resolved measurements, and we chose the BPF set having a center wavelength from 700 to 890 nm, with a 10 nm step.

Furthermore, to selectively excite the RBM CP of one (*n,m*) species of SWCNTs, we generated a train of pump pulses with a repetition rate that coincides with the RBM frequency of the SWCNTs, using the technique of pulse shaping [52, 53]. As schematically shown in Fig. 10, the pump beam enters the pulse shaping setup consisting of a pair of gratings with a patterned mask in between. The different frequency components of the input, single pulse are spatially dispersed by the first grating, and the amplitude and/or phase of each frequency component is modified by the mask. A second lens and grating recombine all the frequencies into a single collimated beam, and then a shaped output pulse train was obtained with the Fourier transformed pulse shape given by the Fourier transform of the pattern transferred by the masks onto the spectrum.

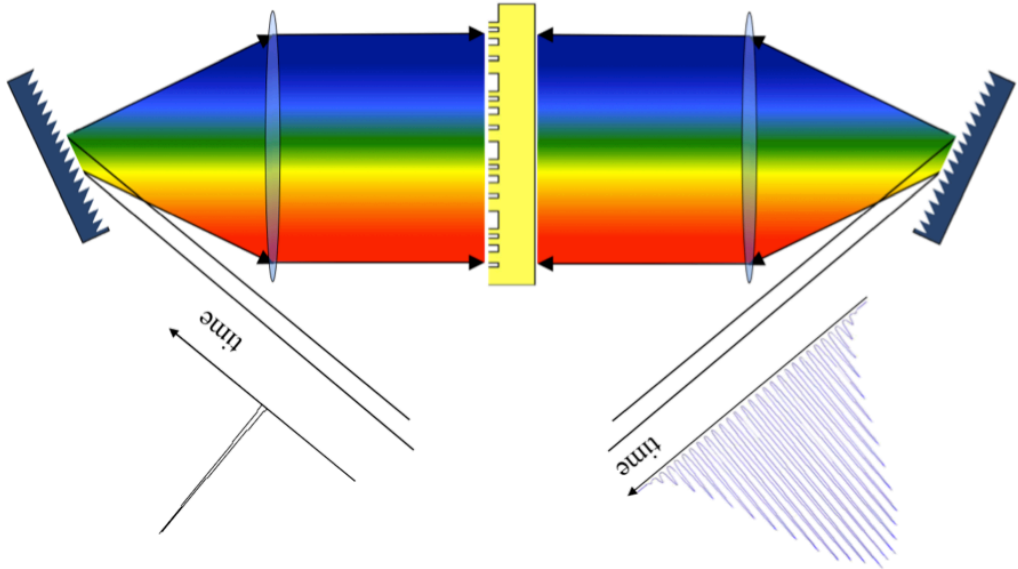

**Figure 10:** Generation of high-repetition-rate pulse trains by spectral filtering, adapted from [53]. (a) input spectrum with the bandwidth of the pulse, $\Delta v$, (b) amplitude filter, (c) phase filter, which has the various shaded rectangles with different phases.

In our experiments, phase masks were fabricated on quartz substrates by standard lithographic techniques with reactive ion etching. This procedure yields a binary phase mask, with the phase difference given by $\Delta\phi = 2\pi(n-1)D/\lambda$, where $n$ is the refractive index, $D$ is the etch depth, and $\lambda$ is the optical wavelength. Pulse trains were generated with these phase masks consisting of periodic repetitions of the *M* (maximal length) sequence [54, 55]. The detail of the mask design are given in Ref. [53]. For a phase retardation of 0.84 $\pi$, the required depth was ~0.73 μm. With a $\pi$-phase delay, the central pulse in the shaped pulse train would be missing from the pulse train [53]. Figure 11 shows the layout of a part of the phase masks we designed for producing a train of pump pulses having a desired repetition rate [56]. This mask contains 25 phase patterns, each of which has a different spatial periodicity and is designed to produce a pulse train with a different repetition rate, ranging from 133 fs (top, 7.5 THz) to 160 fs (bottom, 6.25 THz) in 2-fs (0.05 THz) intervals. The phase masks were sufficiently wide to pass the entire input bandwidth.

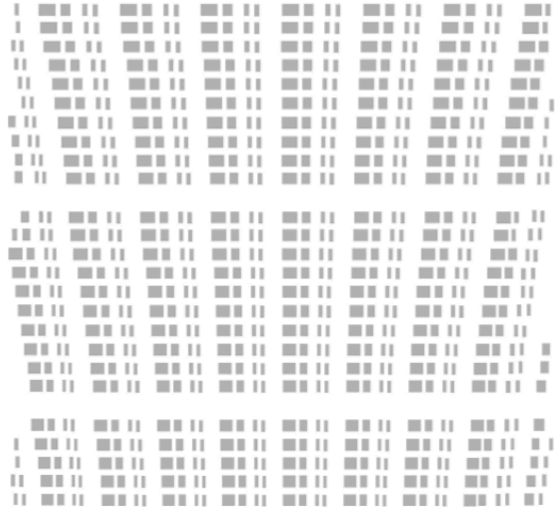

**Figure 11:** Layout of a part of the phase mask designed to produce pulse trains with repetition rates ranging from 133 fs (top, 7.5 THz) to 160 fs (bottom, 6.25 THz) for our specific experiments. The repetition period increases by 2 fs per pattern from top to bottom. The shaded regions are not etched during the fabrication process.

## 4. Experimental results and discussion

*4.1 Real-time observation of coherent lattice vibrations in SWCNTs*

With the advent of ultrafast spectroscopy, one can probe time dependent electronic and vibrational dynamics [16, 57]. In this section, we discuss the time dependent observation of lattice vibrations in SWCNTs. Using pump-probe spectroscopy, we generated and detected the radial breathing mode (RBM) CPs in individual SWCNTs, with no photoluminescence or Rayleigh scattering background, and with excellent resolution as compared with continuous-wave resonant Raman scattering (CW RRS).

The sample used in this study was a micelle suspended SWCNT solution with tube diameters ranging from 0.7-1.3 nm [58]. The tubes were individually suspended with sodium dodecyl sulfate in $D_2O$ via ultrasonication and centrifugation. Using a Ti:sapphire laser with a ~50 fs pulse and ~20 mW average pump power, we performed degenerate pump-probe measurements at room temperature. We tuned the center wavelength in 5-nm steps from 710 to 860 nm (1.75-1.43 eV) by controlling the slit between the intra-cavity prism pair in the Ti:sapphire laser. For comparison, CW RRS experiments were performed on the same sample, for which the excitation source was a CW Ti:sapphire laser with a power of 15 mW, and signal collection was done using a triple monochromator and a CCD camera.

Figure 12(a) shows a typical example of the raw pump-probe data taken on our SWCNT sample at an 800 nm center photon wavelength. The large electronic contribution is seen at time-zero along with a weaker decaying oscillatory component due to the RBM CPs. Figure 12(b) shows CP oscillations in SWCNTs excited at different pump photon energies. Each trace consists of a superposition of multiple RBM modes with different frequencies, exhibiting a strong beating pattern, which changes with the photon energy. The observed CP oscillations are dominated by RBM modes, which are resonantly enhanced by pulses commensurate with their unique electronic transitions. The decay time of the dominant RBM CP oscillations was ~5 ps.

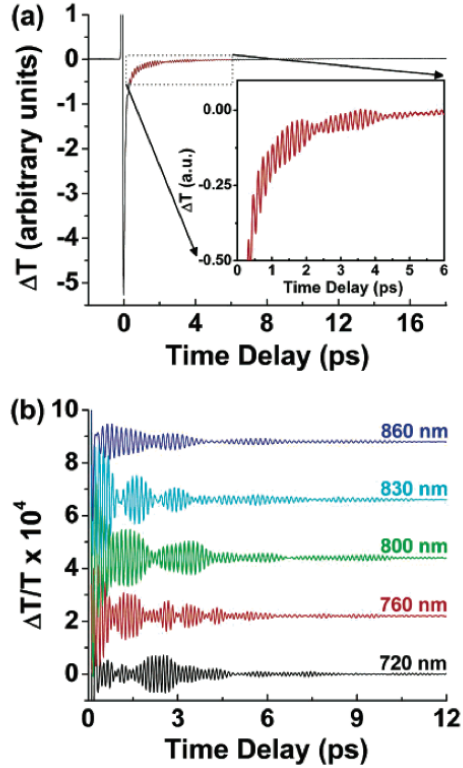

**Figure 12:** (a) Raw pump-probe time-domain data taken at an 800 nm center photon wavelength. (inset) zoom-in of the data between 0.3 and 6 ps. (b) CP oscillations excited and measured at five different photon energies. The traces are offset for clarity. Reproduced from Ref. [1].

To determine the frequencies of the excited lattice vibrations, we took a fast Fourier transform (FFT) of the time-domain oscillations (subtracting off the slow background signal due to the photo-exicted carrier dynamics) shown in Fig. 12(b) and display the results in Fig. 13(a). For comparison, we show CW RRS spectra in Fig. 13(b). Strong CP spectra are seen in three distinct frequency regions, similar to what is seen in CW RRS spectra. The main peak positions in Fig. 13(a) and 13(b) coincide, indicating that the oscillations in Fig. 12(a) are due to RBM CPs. However, we find a few noticeable differences between the CP and CW RRS data. First, unresolved shoulder features in RRS are seen as clear peaks in the CP spectra due to the narrow line widths. Second, there are different intensity distributions among the strongest peaks in the three distinct frequency regions. Finally, the lineshapes in the CP spectra show a surprising double-peaked dependence on the photon energy. (This has been discussed briefly in section 2 above and is related to the modulation of the band gap by the RBMs).

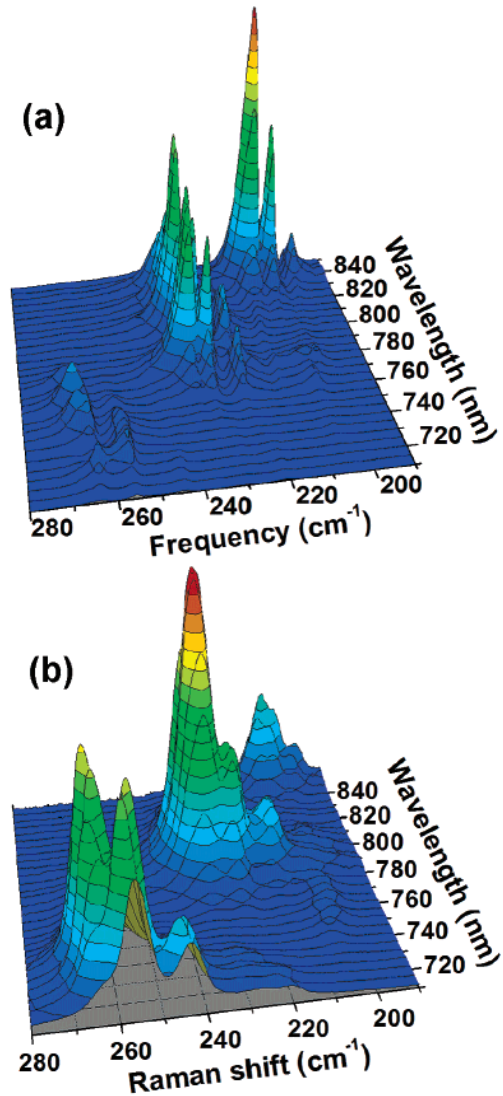

**Figure 13:** (a) A 3D plot of the FFT of CP oscillations obtained over a photon energy range of 710-850 nm (1.75-1.46 eV) with a 5-nm step size. (b) A 3D plot of RRS over an excitation energy range of 710-850 nm (1.75-1.46 eV) with a 5-nm step size. Reproduced from Ref. [1].

For a comparison of the line widths between CP spectra and CW RRS spectra, Figs. 14(a)-(c) are presented. The CP spectra were obtained with different wavelengths are are overlaid with the equivalent RRs spectra taken at the same wavelengths of 710 nm (1.75 eV), 765 nm (1.62 eV), and 830 nm (1.49 eV), respectively. While there is overall agreement between the two sets of spectra in each figure, it is clear that several features are more easily resolved in the CP data. The narrower CP line widths make it possible to resolve blended peaks in the RRS data. With less overlap between nearby peaks, more precise determination of line positions is possible. Through peak fitting to Lorentzian lineshapes, we have identified and assigned 18 RBM peaks in the CP spectra with an average measured line width of ~3 cm$^{-1}$, while line widths measured in CW RRS were consistently 5 – 6 cm$^{-1}$.

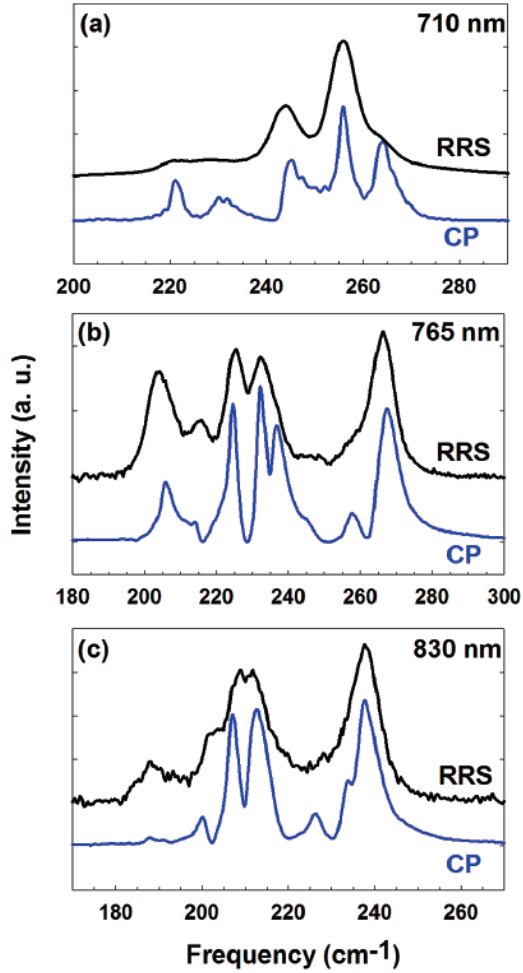

**Figure 14:** Phonon spectra using three different photon energies obtained from RRS and CP spectroscopy measurements. The traces are offset for clarity. Reproduced from Ref. [1].

In order to explain the origin of the double peak, we finally discuss how the generation of coherent RBM phonons modifies the electronic structure of SWCNTs and how this can be detected as temporal oscillations in the transmittance of the probe beam. The RBM CP is a macroscopic vibration of the nanotube lattice in the radial direction so that the diameter ($d_t$) periodically oscillates at an angular frequency $\omega_{RBM}$. The electronic states adiabatically follow the the RBM oscillations. This approximately causes the band gap $E_g$ to oscillate at $\omega_{RBM}$ since $E_g$ is inversely proportional to the nanotube diameter [see Fig. 15(a) and also Fig. 3(a)]. As a result of this bandgap oscillation, the interband transition energies oscillate in time, leading to ultrafast modulation of the optical constants at the angular frequency $\omega_{RBM}$. Furthermore, these modulations also imply that the absorption coefficient at a fixed probe photon energy is modulated at $\omega_{RBM}$. Equivalently, the photon energy dependence of the CP signal shows a derivative-like behavior [see Figs. 15(b)-(c) and Fig. 4]. We modeled this behavior assuming that the CP signal intensity is proportional to the absolute value of the convoluted integral of the first derivative of a Lorentzian absorption line and a Gaussian probe beam profile. The symmetric double-peak feature in Figs. 15(b) and 15(c) confirms the excitonic nature of the absorption line, in contrast to the asymmetric shape expected from the 1D van Hove singularity shown in Fig. 4.

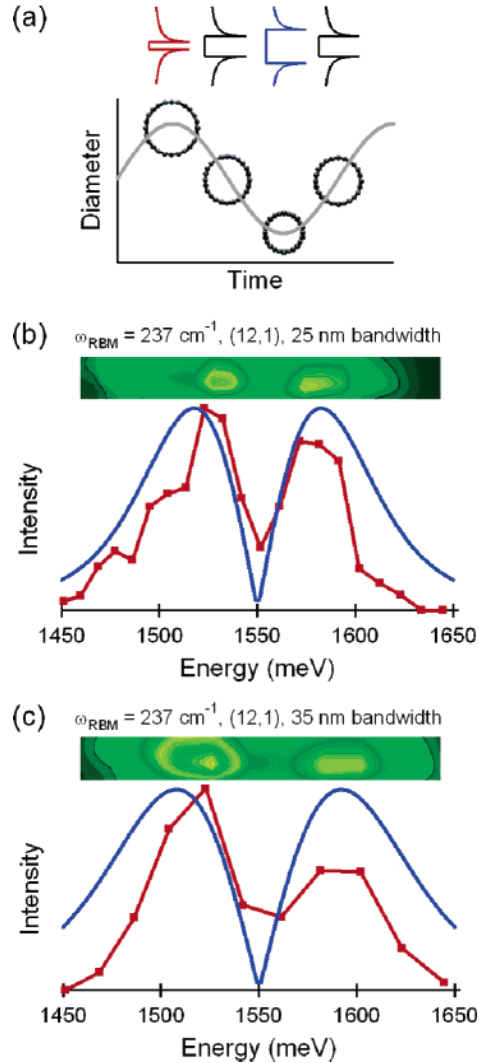

**Figure 15:** (a) Time-dependent band gap due to the RBM of coherent lattice oscillations. (b,c) The photon energy dependence of the CP signal intensity (both contour and 2D plots) for the (12,1) tube with a probe bandwidth of (b) 25 nm and (c) 35 nm, together with theoretical curves. (solid line without dots). Reproduced from Ref. [1].

*4.2 Chirality selective excitation of coherent RBM phonons in SWCNTs by using tailored femtosecond pulse trains*

In this section, a novel method is presented that allows us to study single-chirality nanotubes in ensemble samples contains nanotubes of many different chiralities. Specifically, we have utilized the techniques of femtosecond pulse shaping [52, 53] [59], as described in Section 3.3, in ultrafast pump-probe spectroscopy of SWCNTs to selectively excite the coherent lattice vibrations of the radial breathing mode (RBM) in nanotubes of specific chiralities.

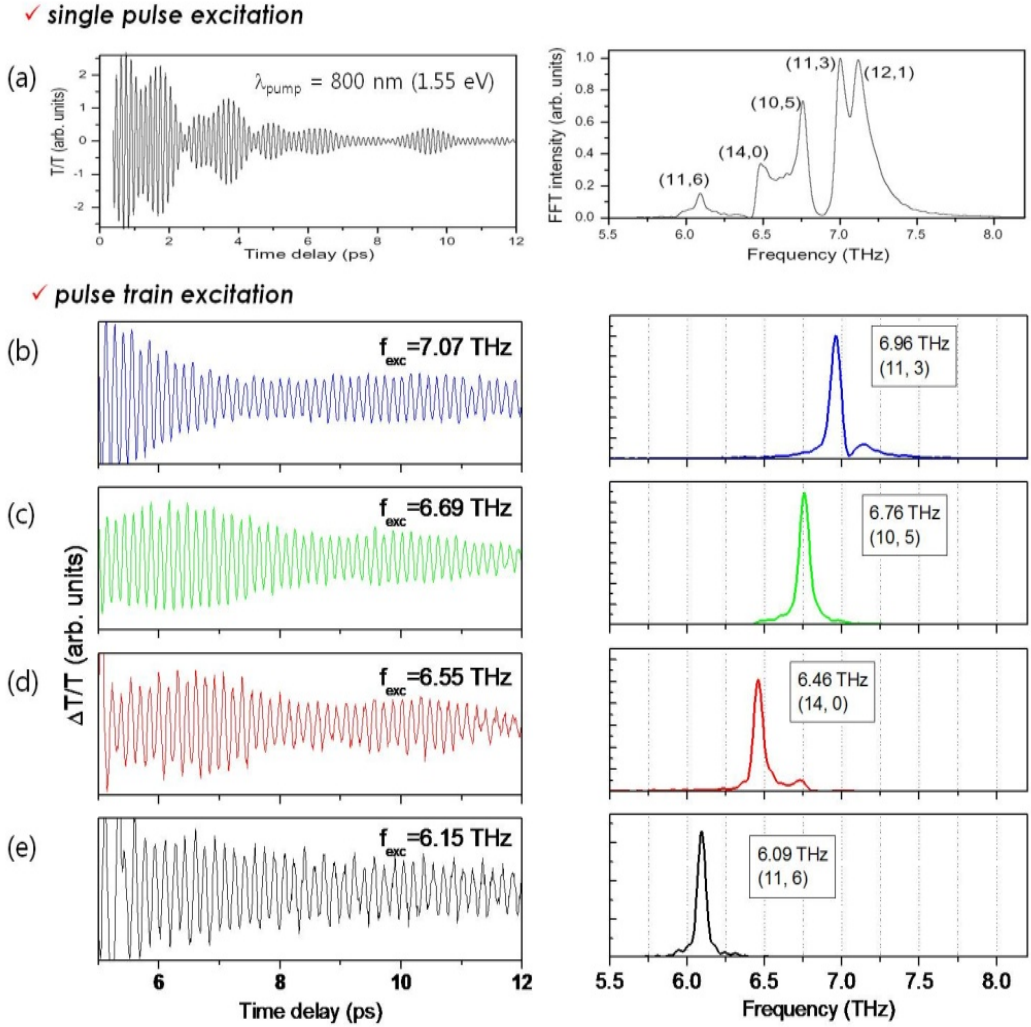

**Figure 16:** (a) (Left):Time-domain transmission modulations of the probe due to coherent RBM vibrations in ensemble SWCNT solution generated using standard pump-probe techniques without pulse shaping, (Right):Fourier transformation of time-domain oscillations with chirality assigned peaks. (b-e) (Left):Time-domain coherent RBM oscillations selectively excited by multiple pulse trains via pulse shaping with corresponding repetition rates of 7.07 THz ~ 6.15 THz, (Right): Fourier transformations of corresponding oscillations, with their dominant nanotube chirality (n,m) indicated. Adapted from Ref. [2].

The optical setup combined pump-probe spectroscopy, with the pulse-shaping apparatus (Fig. 11). The pulse-to-pulse interval in multiple pulse trains was selected to correspond to the period of a specific RBM. Among different species of nanotubes, those having RBM frequencies that are matched to the repetition rate of multiple pulse trains will generate large-amplitude coherent oscillations with increasing oscillatory response to each pulse, while others will have diminished coherent response [56, 60, 61].

The strength of our pulse-shaping technique is shown in Fig. 16. Real-time observation of coherent RBM oscillations is possible without pulse-shaping by employing standard femtosecond pump-probe spectroscopy [1, 45]. Figure 16(a) shows the resulting transmission modulations of the probe beam induced by coherent phonon lattice vibrations, which were generated by pump pulses with pulse duration of 50 fs and a central wavelength of 800 nm. The time-domain beating profiles reflect the simultaneous generation of several RBM frequencies, which are clearly seen in the Fourier-transform of the time-domain data in the right of Fig. 16(a). However, we cannot obtain detailed information on dynamical quantities such as the phonon oscillation phase because the phonon modes overlap one another.

By introducing pulse-shaping, using multiple pulses with different repetition rates to excite RBM CP oscillations, chirality selectivity was successfully obtained as shown in Figs. 16(b)-(e). With the appropriate pulse train repetition rate, a specific chirality RBM CP oscillation dominates the signal while the RBM CP oscillations from other nanotubes are suppressed. For example, by choosing a pump repetition rate of 7.07 THz, we can selectively excite the (11,3) nanotubes, as shown in Fig. 16(b). Similarly, with a pump repetition rate of 6.69 THz, the (10,5) nanotubes are selectively excited [see in Fig. 16(c)]. We note that both nanotubes contributed when the pump repetition rate was tuned to the middle of the RBM frequencies for (11,3) and (10,5) nanotubes [2]. The accuracy of selectivity depends on the number of pulses in the tailored pulse train as well as on the distribution of chiralities in the nanotube ensemble. Also, selective excitation of a specific chirality requires the pump energy to be resonant with the corresponding second interband (or $E_{22}$) transition (see Fig. 17).

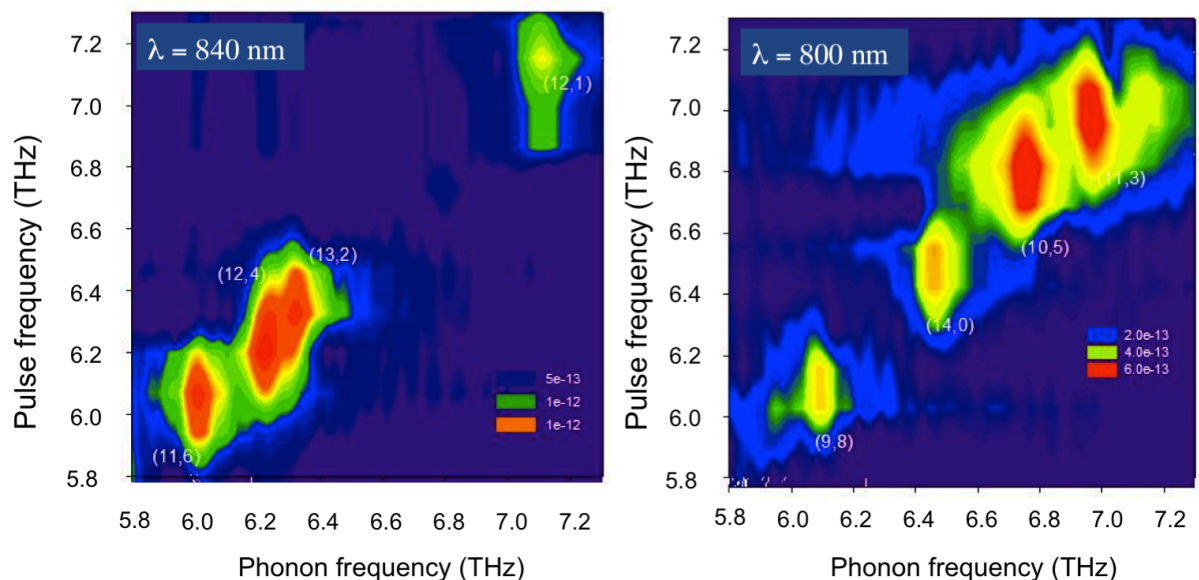

**Figure 17:** Specific chiralities are excited for different resonant excitation energies at (a) 840 nm (1.48 eV) and (b) 800 nm (1.55 eV). In the case of 1.48 eV excitation energy, by choosing a pump repetition rate of 6.1 THz, the (11,6) nanotubes are selectively excited. Similarly, we can selectively excited the (12,1) nanotubes with a pump repetition rate of 7.2 THz.

By placing a series of 10-nm band pass filters in the probe path before the detector, we can measure the wavelength-dependence of RBM-induced transmission changes in order to understand exactly how the tube diameter changes during coherent phonon RBM oscillations and how the diameter change modifies the nanotube band structure. As seen in Fig. 18, the differential transmission is shown for three cases, corresponding to wavelengths below resonance, at resonance, and above resonance, respectively, for selectively-excited (11,3) carbon nanotubes. The amplitude and phase of oscillations vary noticeable for varying probe wavelengths. Specifically, the amplitude of oscillations becomes minimal at resonance, and in addition, there is clearly a $\pi$-phase shift between the above- and below-resonance traces. Because the band gap energy and diameter are inversely related to each other in SWCNTs, and because it is the RBM frequency at which the diameter is oscillating, we can conclude from this data that the energy of the $E_{22}$ resonance is oscillating at the RBM frequency. In other words, when the band gap is decreasing, absorption above (below) resonance is decreasing (increasing), resulting in positive (negative) differential transmission.

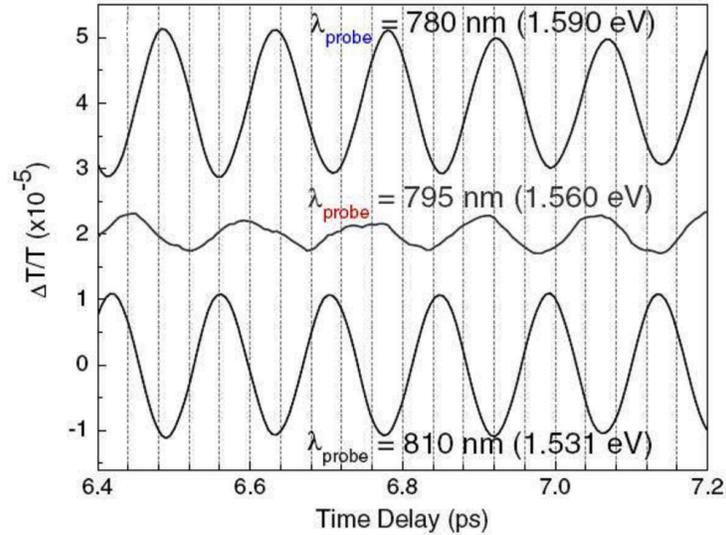

**Figure 18:** Differential transmission as a function of time delay at probe wavelengths of 780 nm, 795 nm, and 810 nm for the selective RBM excitation of the (11,3) nanotubes There is a $\pi$ phase shift between the 780 nm and 810 nm data. These three wavelengths correspond to below, at, and above the second exciton resonance, respectively, of (11,3) nanotubes.

*4.3 Coherent G-mode lattice vibrations in SWCNTs*

In this section, the generation and detection of coherent oscillations of the G-band phonon in SWCNTs with frequency ~ 1580 cm$^{-1}$ (period ~ 20 fs) is discussed [6]. A previous ultrafast optical study of SWCNTs by Gambetta *et al.* revealed anharmonic coupling between the coherent RBM and tangential G-band, leading to a frequency modulation of the G mode by the RBM [45]. Also, Kato and co-workers observed a complex polarization dependence of coherent G-mode oscillations in aligned SWCNTs, which was explained in terms of a superposition of G band phonons with different symmetries [24]. We performed degenerate pump-probe measurements with broad band light source from a Kerr-lens mode-locked Ti:sapphire laser with a pulse duration of about 12 fs. The center wavelength of the laser spectrum was around 800 nm (1.55 eV), with a spectral bandwidth of about 100 nm (200 meV), which is comparable to the G$^+$-mode vibrational energy (197 meV) in SWCNTs. The probe pulses were spectrally filtered after the sample (before the detector), by using a series of band-pass filters with a 10-nm band pass centered at various wavelengths. Pump-induced and spectrally-resolved transmission or scattering modulations were measured as a function of the time delay between pump and probe pulses.
22

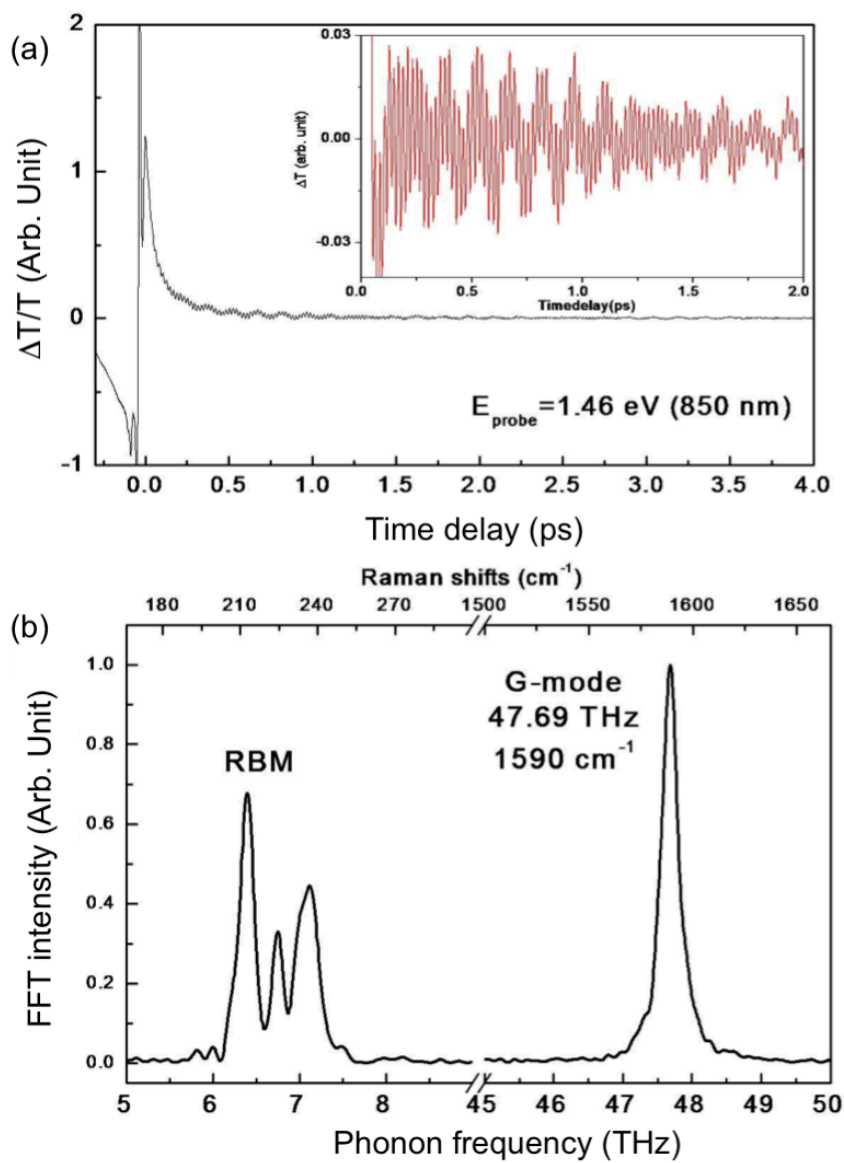

**Figure 19:** Coherent phonon oscillations in single-walled carbon nanotubes and their fast Fourier transform power spectrum are shown. (a) Time-domain probe intensity modulations measured at 850 nm detection wavelength using a Band-pass filter (BPF). The low frequency oscillation with 160 fs is RBM mode, and tangential G-mode emerges through the oscillation of 21 fs period. (b) Corresponding FFT spectrum in the frequency domain showing RBMs from 6.0 ~ 7.5 THz (200~250 cm-1) and G-mode at 47.69 THz (1590 cm$^{-1}$). Adapted from Ref. [6].

Figure 19(a) shows the raw pump-probe experimental results in isolated SWCNTs. Time-domain intensity modulations of a probe beam of energy ~ 1.46 eV (850 nm) are clearly seen. The trace in the inset was obtained after subtracting the overall exponential decaying components that correspond to the relaxation process of photo-excited electrons [62]. The oscillatory signal, which originates from the coherent lattice vibrations excited by the pump pulse, is composed of high-frequency and low-frequency contributions. As is confirmed from the fast Fourier-transformed power spectrum in Fig. 19(b), the low-frequency signal at around 7 THz, corresponding to RBM peaks, reveals that the sample contains several chiralities of SWCNTs, each having a different RBM frequency, that are resonantly excited at 850 nm [1]. The focus in this section is centered on the dynamics and the detection mechanism for the high frequency

G-mode oscillations of SWCNTs, having a vibrational frequency of 47.69 THz (1590 cm$^{-1}$) in the Fourier-transformed spectrum.

As fully described in Sections 2.1.4 and 4.1, for the RBM CP oscillations, the change in the absorption coefficient due to RBM-induced bandgap oscillations provide a straightforward explanation for the appearance of CP oscillations in pump-probe differential transmission. The G-mode distortion of the lattice can also modify optical constants of carbon nanotubes, but our theoretical calculations [7] suggest that the absorption coefficient modulations due to the G-mode vibrations are expected to be 1000 times smaller than those due to the RBM. Thus, from the fact that the coherent G-mode signal is comparable to the RBM signal as in Fig. 19(b) as well as from the dependence of the signal on the probe energy, we came to a conclusion that a different detection mechanism is at play for the G-mode.

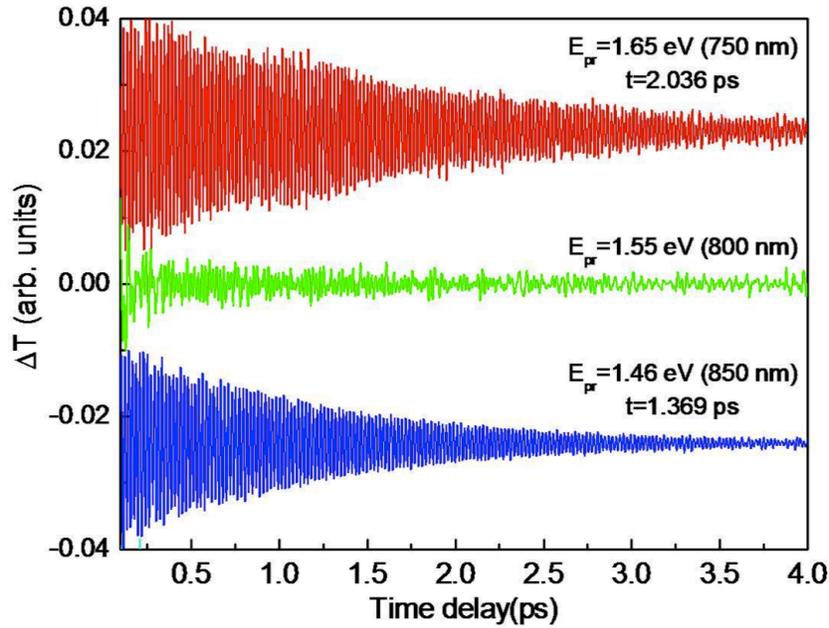

**Figure 20:** Coherent G-mode oscillations in the time-domain are measured at different probe energies, 1.65 eV (750 nm), 1.55 eV (800 nm), and 1.46 eV (850 nm) using band-pass Filters depicted as the upper scheme. From these oscillations, the phonon decay times of the G-mode oscillations are obtainable about 2.036 ps at the 1.65 eV detection energy and 1.369 ps at the 1.46 eV energy. Adapted from Ref. [6].

As the spectral window for the probe pulse is applied by band-pass filters depicted in the upper part of Fig. 20, the G-mode oscillations exhibit a strong dependence on the probe wavelength in amplitude. The lower part of Fig. 20 shows the probe intensity modulations from the coherent G-mode oscillations at different probe energies, 1.65 eV (750 nm), 1.55 eV (800 nm), and 1.46 eV (850 nm). It is interesting to see that that the signal is almost completely suppressed when the probe signal energy is close to the center of the laser spectrum, while strong oscillations are observed at 1.65 eV or 1.46 eV, which are each separated from the center energy by roughly half of the G-mode energy (~100 meV). Additionally, the G-mode amplitude decays monotonically with time delay for each wavelength while having slightly different dephasing times and the signal at lower probe energy tends to decay faster than that at higher energy.

Figure 21(a) shows the G-mode amplitude as a function of the probe energy. The amplitude curve features two peaks with each maximum occurring near the probe energy of 1.46 and 1.66 eV, respectively, which are separated from each other by the G-mode energy, while having a local minimum near the center energy of the laser spectrum. We show below that this intriguing probe wavelength dependence of the amplitude shown in Fig. 21(a) can be explained by taking the coherent Raman processes into account.

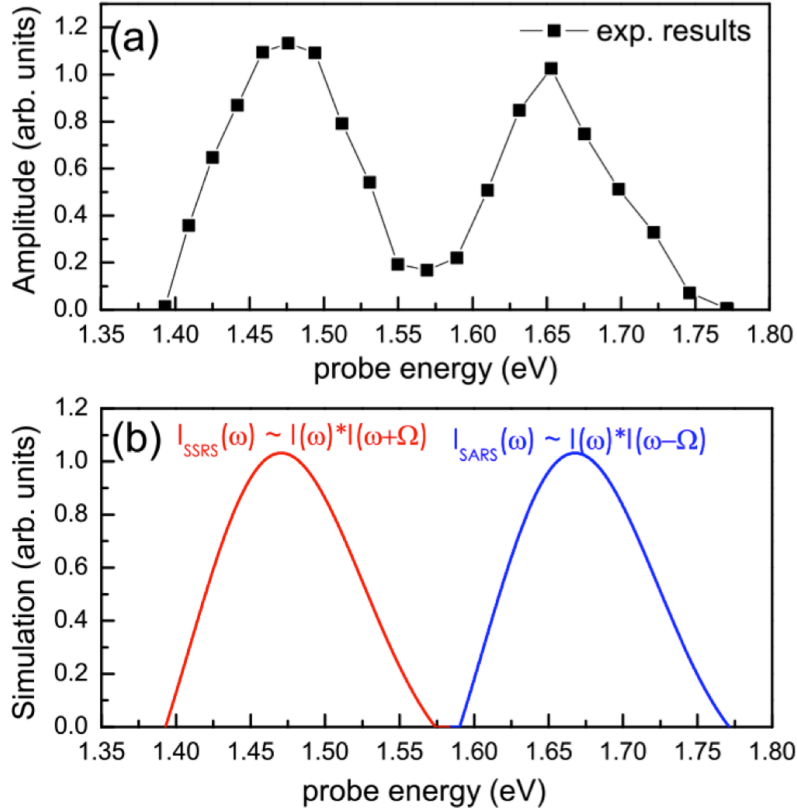

**Figure 21:** (a) G-mode oscillation amplitude as a function of the probe energy, featuring two peaks located near 1.47 eV and 1.66 eV, which correspond to the signal from the SSRS and the SARS processes (see Fig. 4.4), respectively. (b) Simulated spectral intensity curves for the SSRS and the SARS processes obtained for a Gaussian laser spectrum centered at 1.55 eV with a spectral width of 195 meV FWHM. Adapted from Ref. [6].

The probability $P$ for stimulated Stokes (anti-Stokes) scattering to occur at a certain energy $E$, will be proportional to both the stimulating photon intensity $I(E)$ and the stimulated Stokes (anti-Stokes) photon intensity $I(E-\hbar\omega)$ ($I(E+\hbar\omega)$) such that;

$$P_{SSRS}(E) \propto I(E) \times I(E-\hbar\omega_G) \times n_G \quad (4.1)$$

$$P_{SARS}(E) \propto I(E) \times I(E+\hbar\omega_G) \times n_G \quad (4.2)$$

where $n_G$ is the occupation number for the coherent G-mode which was generated by the pump pulse and $\hbar\omega_G$ the G-mode phonon energy. Simulations of the scattering intensity for the two coherent Raman scattering processes, as shown in Fig. 21(b), were carried out assuming a Gaussian laser spectrum centered at 1.55 eV with a FWHM spectral width of 195 meV, similar to our experimental conditions. The relatively good agreement between the experimental results in Fig. 21(a) and the simulations imply that the detection mechanism is due to impulsively stimulated Stokes and anti-Stokes Raman processes in the probe pulses. Thus, we assign the coherent G-mode signal measured near 1.46 eV to the contribution from the SSRS process for incoming photons near 1.66 eV, and the signals near 1.66 eV are generated through the SARS process from the incoming photons near 1.46 eV. We note that the energy of the amplitude dip changed accordingly as we tuned the center of the laser spectrum.

For impulsively stimulated Raman scattering, where two photons whose energies are separated by the phonon energy are incorporated, the time sequence between those photons may influence the scattering efficiency, especially when the real electronic transitions of the nanotubes are incorporated for the stimulated photons, such that the excited state can be sustained for an extended period. If a photon with a higher energy precedes a lower energy photon, the SSRS process will be more efficient than the SARS process, and vice versa.

The time sequence of the photons can be controlled by adding or subtracting dispersion by adjusting the optical path length through the fused silica prism, which was used initially for dispersion compensation of all the optical elements to get the shortest pulse at the sample position. The modification of the dispersion can result in chirp such that the high (low) wavelength components arrive earlier than the low (high) components for the case of negative (positive) dispersion, shown in Fig. 22.

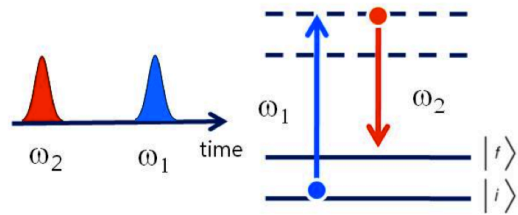

Negative dispersion

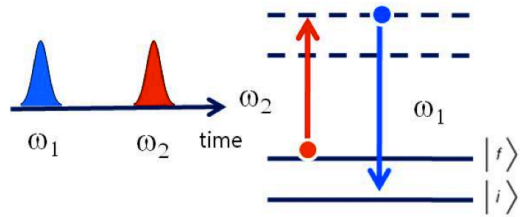

Positive dispersion

**Figure 22:** Scheme of time sequence of photons by controlling the dispersion. The modification of the dispersion can result in chirp such that the high (low) energy components arrive earlier than the low (high) components for the case of negative (positive) dispersion. Adapted from Ref. [6].

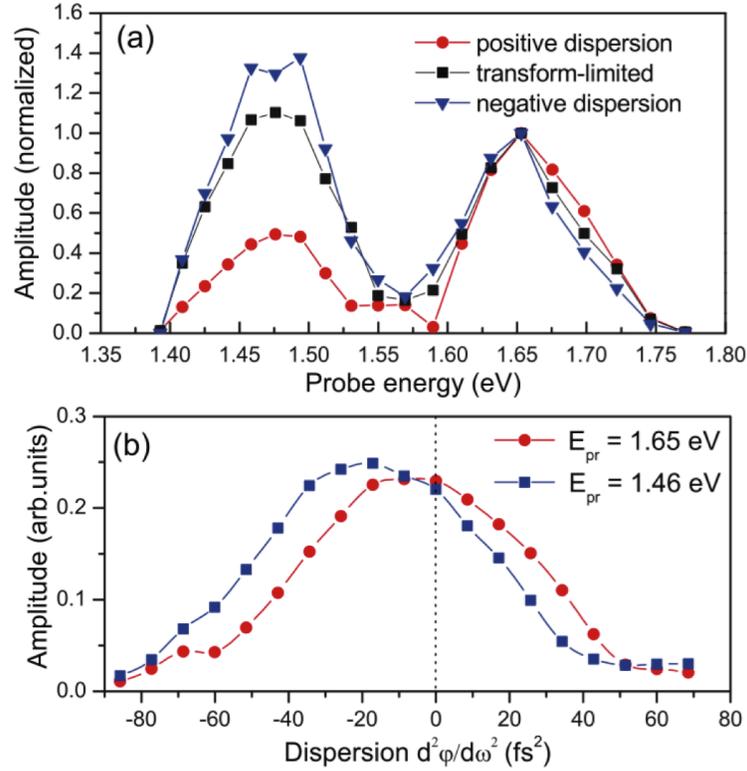

**Figure 23:** (a) Oscillation amplitudes as a function of probe wavelength, which were obtained using negative dispersion (red), transform-limited (black), and positive dispersion (blue) pulse chirps. (b) Phonon amplitude at 1.66 eV corresponding to the SARS contribution and that at 1.47 eV corresponding to the SSRS contribution, measured as a function of applied dispersion. Adapted from Ref. [6].

By controlling the dispersion in ultrashort pulse, we preferentially drive one process between SSRS and SARS. The SARS (SSRS) process is dominant on the positive (negative) dispersion, as shown in Fig. 23(a). Figure 23(b) shows the phonon amplitude at 1.65 eV corresponding to the SARS contribution and that at 1.46 eV of the SSRS one as a function of the applied dispersion value. For positive (negative) dispersion, the SARS (SSRS) signal is stronger than the SSRS (SARS) signal, which is in good agreement with the consideration of the order between stimulating and stimulated photons for the detection processes. We note that the dispersion given to the pump pulse can modify the excited phonon amplitude but cannot explain the observed dependence of each coherent Raman process occurring within probe pulses.

*4.4 $E_{11}$-resonant coherent phonon generation in SWCNTs*

In Section 4.1, we described the observation of coherent phonon (CP) oscillations of radial breathing modes in SWCNTs generated via impulsive excitation of $E_{22}$ optical transitions [1, 2, 7]. We also showed CP spectroscopy has several advantages over Raman spectroscopy, including no Rayleigh scattering and PL backgrounds. In this section, we use these advantages to study CP oscillations of RBMs in smaller-diameter SWCNTs, synthesized by the CoMoCAT method, which have $E_{11}$ transitions within the wavelength range accessible with a Ti:sapphire laser.

The sample used in this study was an aqueous suspension of CoMoCAT SWCNTs [4]. Multiple RBMs of SWCNTs corresponding to different diameters were simultaneously excited within the broad 30-40 nm bandwidth of 40-50 fs pulses from a Ti:sapphire oscillator through degenerate pump-probe differential transmission spectroscopy. We tuned the center wavelength of the pump beam over a wide wavelength range of 720 – 1000 nm in steps of 5 nm to investigate the $E_{11}$ and $E_{22}$ transitions.

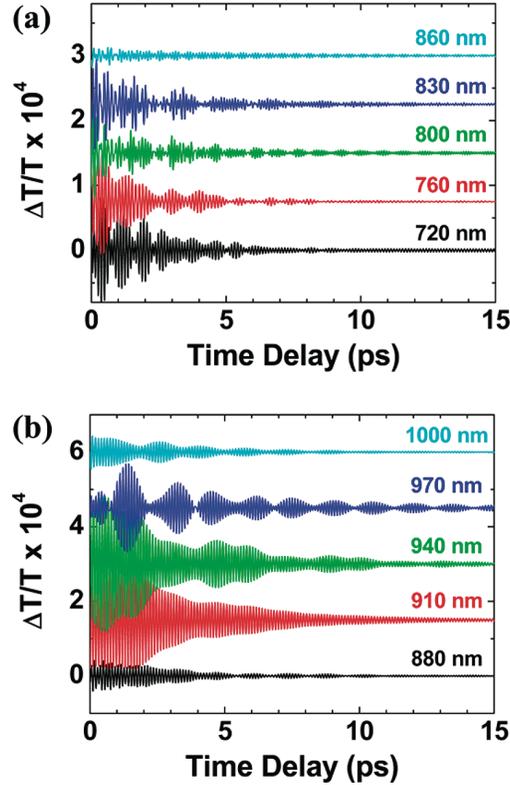

**Figure 24:** Coherent phonon oscillations measured at center wavelengths of (a) 720 nm to 860 nm and (b) 880 nm to 1000 nm, using degenerate pump and probe pulses. Reproduced from [4].

Figure 24 shows CP oscillations of RBMs resonantly excited through $E_{11}$ and $E_{22}$ optical transitions at selected excitation wavelengths within the 720 – 1000 nm (1.23 eV – 1.71 eV) range. Each trace shows a strong beating pattern due to the simultaneous excitation of multiple RBMs, which sensitively changes with the photon energy, implying that the CP oscillations are dominated by a few resonantly excited RMBs. The data at the longer wavelengths [see Fig. 24(b)] show a normalized differential transmission ($\Delta T/T$) of the order of ~$10^{-4}$ near time zero, which is 2 - 3 times larger than that for the shorter wavelength excitation [Fig. 24(a)].

Figure 25 (a) shows contour plot of the CP intensity on a log scale as a function of photon energy (1.23-1.71 eV) and RBM frequency (155-400 cm$^{-1}$), obtained through a Linear Predictive Singular Value Decomposition (LPSVD) analysis on the measured time-domain data [4]. Figures 25(b) and 25(c) show representative CP spectra for excitation wavelengths of 720-740 nm ($E_{22}$ optical transition) with a step size of 5 nm, exhibiting three dominant RBMs between 240 and 280 cm$^{-1}$, corresponding to (11,0), (10,2), (9,4), and (8,6) SWCNTs. The peak at 306 cm$^{-1}$ is due to the $E_{22}$-excitaed RBM of (9,1) tubes ($E_{22} \sim 1.76$ eV), with some contribution form (6,5) tubes excited between $E_{11}$ and $E_{22}$. The black curve in Figure 4.4.2 (c) was taken with 765 nm (1.62 eV) excitation, showing $E_{22}$-resonant CPs for the $(2n + m) = 25$ family [(12,1), (11,3), and (10,5)].

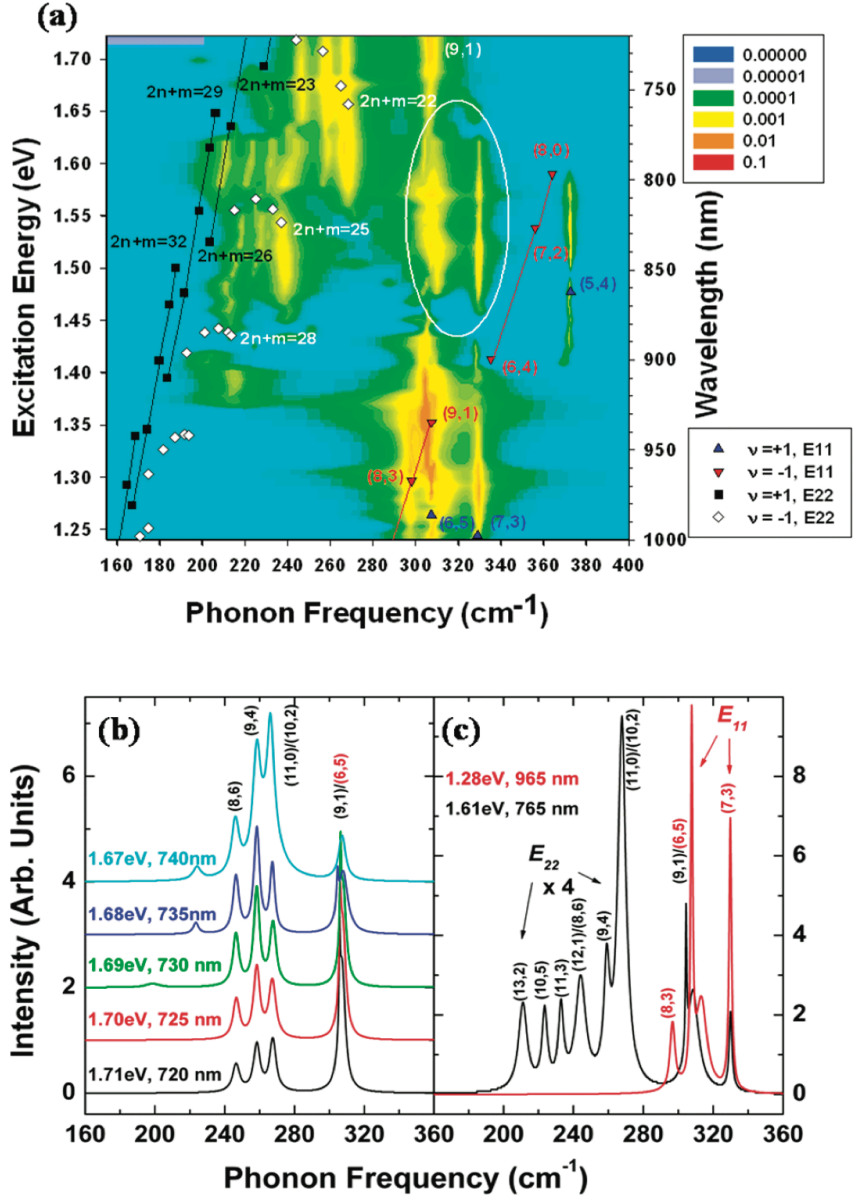

**Figure 25:** (a) Two-dimensional log plot of a Fourier transform of CP oscillations measured over a wavelength range of 720 – 1000 nm (1.71 – 1.23 eV) with a 5 nm step size. The squares are for $E_{22}$ transitions, and triangles are for $E_{11}$ transition from Ref. [63]. Here, $v \equiv \mod 3(n-m)$. (b) FFT power spectra for CP oscillations measured at center wavelengths of 720, 725, 730, 735, and 740 nm, showing RBMs resonantly excited through $E_{22}$ optical transitions. (c) Comparison of RBMs resonantly excited through $E_{22}$ and $E_{11}$ optical transitions with degenerate pump-probe pulses with center wavelengths at 765 and 965 nm, respectively. Reproduced from Ref. [4].

*4.5 Polarization dependence of coherent phonon in highly aligned SWCNTs*

Given the quasi-1D nature of SWCNTs, single-tube and ensemble samples have strong polarization anisotropy that is dominant in photoluminescence (PL), absorption, and Raman measurement [64-69]. Similar anisotropy effects are expected with bulk samples [70]. Here, the anisotropic optical and vibrational properties of a bulk film of highly-aligned SWCNTs are studied using polarization dependent coherent phonon (CP) spectroscopy and a microscopic theoretical model is developed to explain the observed extreme anisotropy [5, 24].

Such highly-aligned bulk films are grown via chemical vapor deposition (CVD), where vertically aligned patterned arrays of SWCNTs are transferred to a sapphire substrates and converted to horizontally aligned SWCNTs having ultralong lengths of ~50 μm and separated by a pitch of ~50 μm. The bulk sample formed using this technique is highly aligned with a diameter distribution ranging between 1 to 5 nm [5, 70, 71].

A mode-locked Ti:sapphire laser was tuned to a central wavelength of 850 nm to excite the $E_{44}$ interband transitions of SWCNTs in this sample via degenerate pump-probe spectroscopy. Two types of polarization-dependent measurements are investigated, as depicted in Fig. 26. Type I configuration maintains the same polarization for the pump and probe, while the alignment axis of the sample is rotated. The Type II configuration maintains the same polarization for the probe and sample, while the pump polarization is rotated by a half-wave plate.

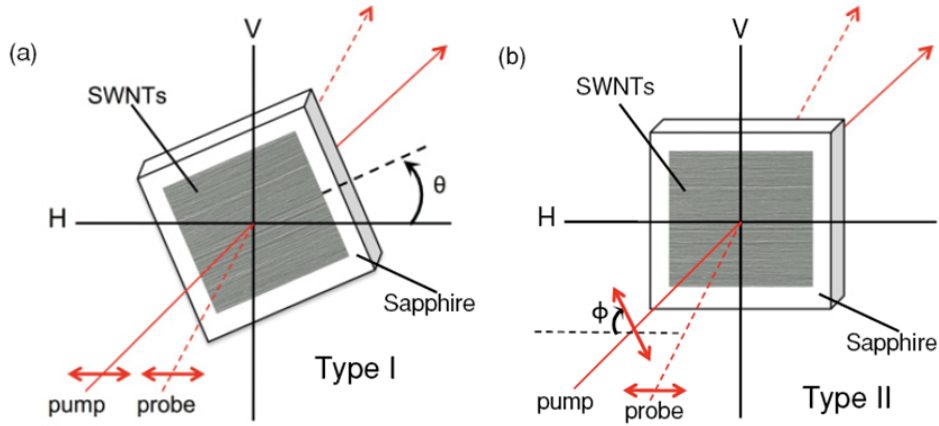

**Figure 26:** Polarization-dependent pump-probe spectroscopy of highly aligned single-wall carbon nanotubes. (a) Type I configuration, where the pump and probe polarizations are fixed and the sample orientation is rotated. (b) Type II configuration, where the probe and sample orientations are fixed and the pump polarization is rotated. Reproduced from Ref. [5].

The results of both Type I and Type II measurements are shown in Fig. 27. As the polarization angle is rotated, a strong polarization anisotropy of the RBM CPs is seen as a function of angle, where the CP signal is almost completely quenched at 90 degrees for both Type I and Type II.

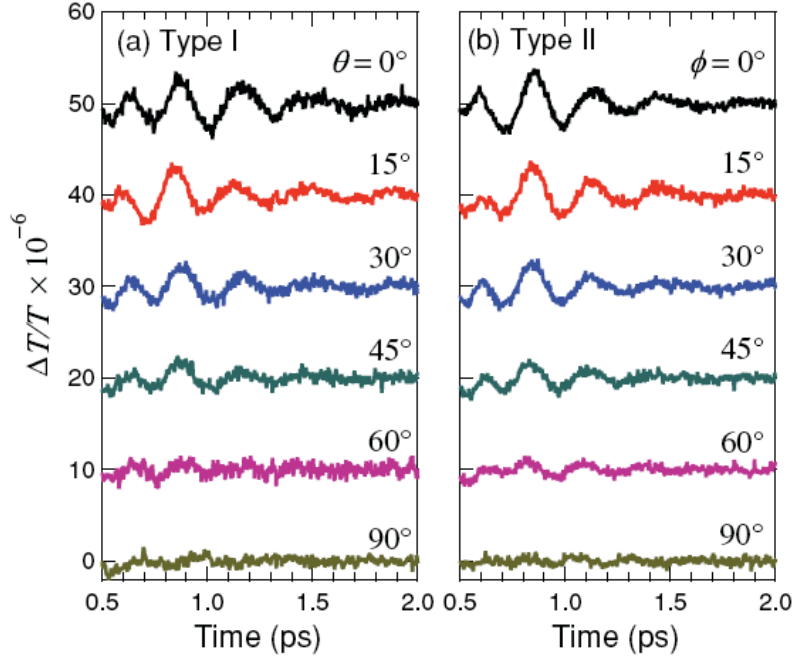

**Figure 27:** Experimental differential transmission of coherent RBM oscillations in highly aligned single-wall carbon nanotubes for different polarization angles in (a) Type I and (b) Type II configurations (see Fig, 26). Reproduced from Ref. [7].

A detailed microscopic theory for the generation and detection of coherent phonons in SWCNTs has been developed that takes into account a realistic band structure and imperfect nanotube alignment in the sample [5, 7]. Figure 28 compares a theoretically determined $\cos^8(\theta)$ dependence for Type I and $\cos^4(\phi)$ dependence for Type II to the experimental results and takes into account the effects of nanotube misalignment. Such a comparison determined the nematic parameter of the sample to be 0.81, which indicates the sample is in fact strongly aligned.

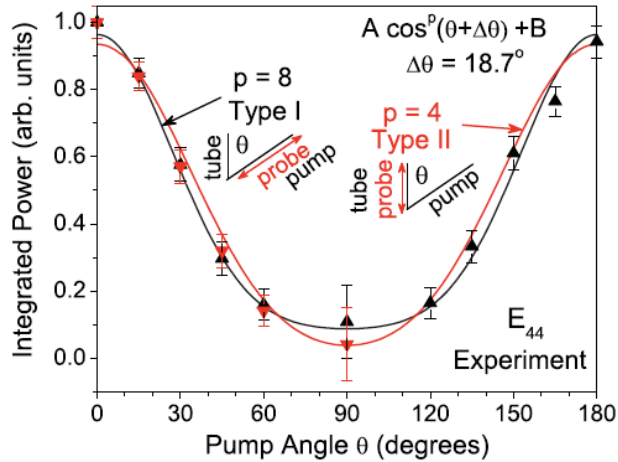

**Figure 28:** Experimental integrated CP power as a function of theta for Type I (black triangles) and phi for Type II (red triangles) orientations. Type I experiment is fit to $A\cos^8(\theta + \Delta\theta) + B$, and Type II is fit to $A\cos^8(\phi + \Delta\phi) + B$, where $A$ and $B$ are background subtraction and rescaling parameters, and standard deviations for the random tube axis misalignment $\Delta\theta$ and $\Delta\phi$ are restricted to the same misalignment parameter for both Type I and Type II. Reproduced from [7].

*4.6 Observation of coherent lattice vibrations in graphene*

Investigations of lattice vibrations in graphitic crystals have been quite fruitful [51]. Raman scattering studies have revealed a correlation between the G-mode frequency in Raman spectra and the number of graphene layers [72], and the spatial mapping of the layer number can be performed by using a scanning confocal Raman spectroscopy [40]. In addition, coherent lattice vibrations in graphite corresponding to an interlayer shearing phonon mode have been observed [73], and the influence of a non-equilibrium electron-hole plasma on the femtosecond dynamics of the in-plane $E_{2g}$ coherent G-mode phonon in graphite has been noted [74]. Here, our recent work on coherent lattice vibrations in graphene films is briefly described [75].

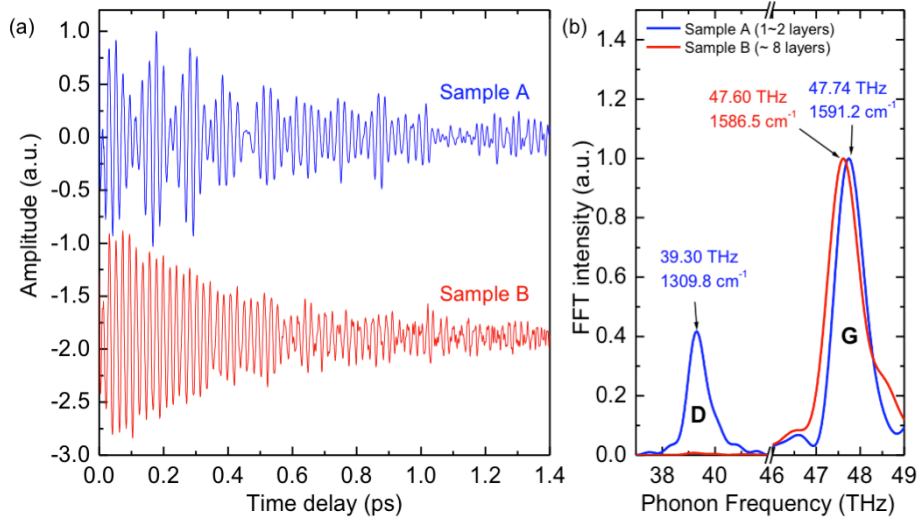

**Figure 29:** (a) Time-domain transmitted intensity modulations of single-layer (upper trace) and multi-layer (lower trace) graphene films. (b) FFT spectra of (a). The G-peak frequency of 47.46 THz (1582 cm$^{-1}$) obtained for the single-layer sample is higher than that of the multi-layer sample at 47.31 THz (1577 cm$^{-1}$). A strong D-peak is observed only for the single-layer sample at 39.21 THz (1307 cm$^{-1}$).

Figure 29(a) shows time-resolved modulation of the probe intensity for a single-layer graphene sample (upper trace) and a multi-layer graphene sample (lower trace), which were obtained after subtracting a slowly decaying component related to the energy relaxation process of photo-excited electrons. The vibrational modes obtained here are the G-band modes at around 1582 cm$^{-1}$ (47.46 THz), which originates from the doubly degenerate zone-centered $E_{2g}$ mode, and the D-band at about 1307 cm$^{-1}$ (39.21 THz), related to defects in graphitic materials. To determine the frequencies of the coherent phonon oscillations, a fast Fourier transform (FFT) of the time-domain oscillations shown in Fig. 29(a) was performed; the resulting FFT spectra, shown in Fig. 29(b), reveal two phonon peaks, the G- and D-peaks. As was revealed through Raman spectroscopy [40, 72, 76, 77], we also find that the position of the G-peak is dependent on the number of graphene layers. The G-peak frequency (1582 cm$^{-1}$) for the single-layer sample is higher than that of the multilayered sample at 1577 cm$^{-1}$ (47. 31 THz). The frequency shift of ~5 cm$^{-1}$ between the two samples, which is considered to originate from the interlayer interaction, is in a good agreement with the results of Raman spectroscopy [72].

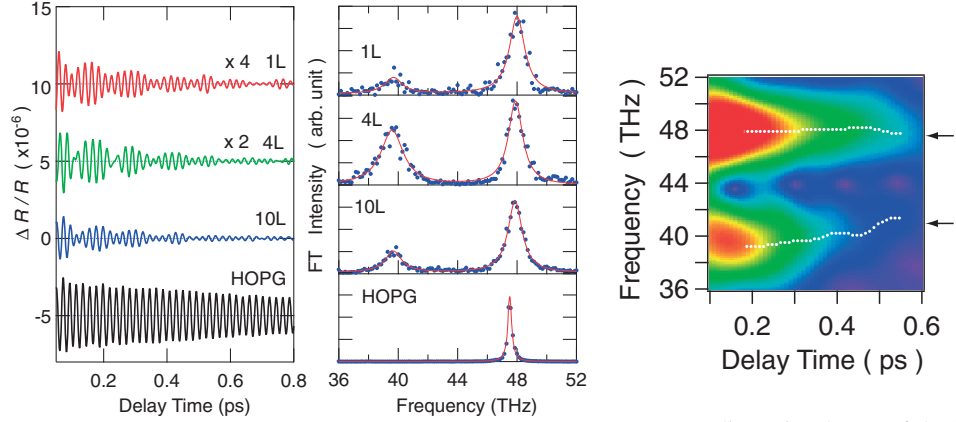

**Figure 30:** (left) Transient reflectivity changes in the high-frequency coherent vibrations in graphene on silicon with different layer numbers and HOPG. (middle) Fourier transform of the time-domain traces in (left). The red curves are Lorentzian fits. (right) Time-frequency map of the high-frequency coherent phonons in four-layer graphene. The dashed lines show the center frequencies of the G- and D-modes, while the arrows indicate the stationary values of the G- and D-modes. Adapted from Ref. [25].

Koga *et al*. has recently observed coherent phonons in graphene on silicon with different numbers of layers [25]. High-frequency coherent vibrations both the G-mode and the D-mode were observed. The G-mode frequency was found to shift to higher frequencies as the number of graphene layers decreased, whereas the D-mode did not exhibit any shifts, as shown in Fig. 30. Moreover, the D-mode was observed to shift to higher frequencies as a function of time. Figure 30(left) shows the transient reflectivity changes in graphene on silicon with different numbers of layers. The changes in a highly oriented pyrolytic graphite (HOPG) are also shown for comparison. The HOPG shows only one oscillatory. component with a long dephasing time (~1.0 ps), while the graphene samples exhibit an apparent modulation of the oscillatory components with a short dephasing time (<0.5 ps). In the Fourier transform [see Fig. 30(middle)] of the data exhibit two peaks located at ~47.5 and ~39.5 THz, corresponding to the G-mode and D-mode, respectively. Figure 30(right) shows the time-frequency two-dimensional map of the two modes. The G-mode frequency is constant in time, whereas the D-mode shifts to higher frequencies with time after 0.2 ps. The authors attribute this shift to photo-excited, carrier-induced bond softening, combined with the highly dispersive nature of the D-mode due to a *k*-selective double resonance process, but more detailed theory is needed to explain the observation.

## 5. Calculations of coherent nanotube RBM and nanoribbon RBLM phonon amplitudes

A special feature found in the RBM CP phenomena in SWCNTs is that the tube diameter can initially expand or contract depending on the nanotube family and laser excitation energy. A similar behavior is also expected in graphene nanoribbons (GNRs), in which we have a radial breathing like mode (RBLM) in the direction of the ribbon width. To understand the physical origin of the phenomena, we calculate the coherent phonon amplitudes using both an extended tight binding (ETB) method and an effective mass theory. We find that the initial expansion and contraction is determined by the *k*-dependent electron-phonon interaction [9, 10]. This results in a family dependence of the CP amplitudes in SWCNTs and armchair GNRs (Sections. 5.1 and 5.2.2, respectively). In the case of zigzag GNRs, the edge states play an important role in the generation of coherent phonons so that the CP amplitudes do not depend on the ribbon size (Section 5.2.1).

## 5.1 Coherent RBM phonons in SWCNTs

Coherent phonon amplitudes are calculated using a computer program (based on a microscopic theory of coherent phonon oscillations) developed by Sanders *et al.*, which solves a driven harmonic oscillator equation derived from the Heisenberg equations of motion [7]. The program makes use of electronic energies and wave functions obtained from an ETB calculation [78], the phonon dispersion relations and corresponding phonon modes [79], the electron-phonon interaction matrix elements [80], the optical matrix elements [81], and the interaction of carriers with an ultrafast laser pulse.

For the first order process of electron-phonon interaction such as RBM (or G) band, only $q = 0$ phonon modes are coherently excited. As we have mentioned many times, for coherent phonons to be excited, it is necessary for the pump pulse to have a duration shorter than the phonon period so that the pump pulse power spectrum has a Fourier component at the phonon frequency. In a simple forced oscillator model neglecting oscillation decays, the coherent RBM phonon amplitude $Q(t)$ with frequency $\omega$ satisfies a driven oscillator equation [7]:

$$\frac{\partial^2 Q(t)}{\partial t^2} + \omega^2 Q(t) = S(t), \tag{5.1}$$

subject to the initial conditions $Q(0) = 0$ and $\dot{Q}(0) = 0$. Here $S(t)$ is a driving force which dependes on the photoexcited carrier distribution function [7]:

$$S(t) = -\frac{2\omega}{\hbar} \sum_{\mu k} M_{\text{el-ph}}^{\mu}(k) \delta f^{\mu}(k,t), \tag{5.2}$$

and $M_{\text{el-ph}}^{\mu}$ is the $k$-dependent RBM electron-phonon matrix element for the $\mu$-th cutting line, and $\delta f^{\mu}$ is the net photogenerated electron distribution function with a pump pulse pumping at the $E_{ii}$ transition energy as obtained by solving a Boltzmann equation for the photogeneration process. The photogeneration rate in the Boltzmann equation depends on the excitation laser energy and also contains the electron-photon matrix element $M_{\text{OP}}$ for light polarized along the tube axis, so that we have the proportionality

$$\delta f^{\mu} \propto M_{\text{op}}^{\mu}. \tag{5.3}$$

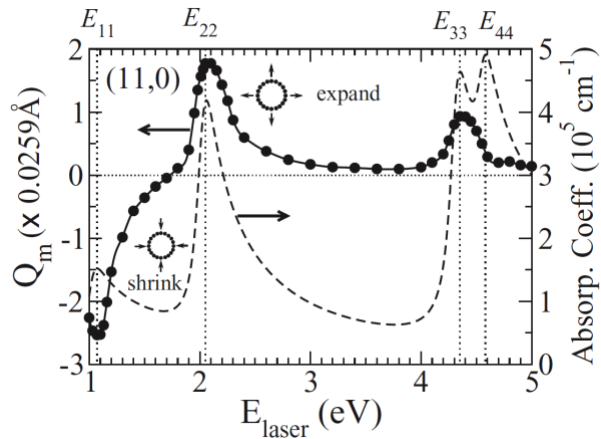

**Figure 31:** The coherent RBM phonon amplitude $Q_m$ (dots) for an (11,0) zigzag tube as a function of laser excitation energy $E_{\text{laser}}$. For clarity, $Q_m$ is plotted in units of 0.0259 Å. A positive (negative) sign of the vibration amplitude denotes a vibration whose initial phase corresponds to an expanding (shrinking) diameter. The absorption coefficient (dashed line) versus $E_{\text{laser}}$ is shown for comparison.

In Fig. 31, we plot the coherent RBM phonon amplitude $Q_m$ (dots) and the absorption coefficient (dashed line) for an (11,0) nanotube as a function of $E_{laser}$. Here $Q_m$ can be defined by taking the vibration of the coherent phonon amplitude to be $Q(t) = Q_m \cos \omega t$, where the origin of time $t = 0$ is given by the first maximum (minimum) of $Q(t)$ and $Q_m > 0$ and $Q_m < 0$ correspond to an initial tube diameter expansion and contraction, respectively. From Fig. 31, we see that the pump light is strongly absorbed at the $E_{ii}$ energies where many photoexcited carriers give rise to a large driving function $S(t)$ in Eq. (5.2) and thus an enhanced $Q_m$. As shown in Fig. 31, $Q_m$ has a negative sign at $E_{11}$, indicating that the tube diameter initially shrinks and oscillates about a smaller diameter than the original diameter for $E_{laser} = E_{11}$, while at $E_{22}$ and higher energies (e.g., $E_{33}$ or $E_{44}$) $Q_m > 0$, that is, the tube diameter initially expands and oscillates about a larger diameter. In general, according to the Franck-Condon principle, an electron is excited without changing the coordinate of the lattice. When a photoexcited electron occupies an anti-bonding electronic state, the adiabatic potential minimum for the ground state is no longer a minimum for the excited state and thus the lattice starts to vibrate around a new adiabatic potential minimum for the photoexcited anti-bonding states. This minimum energy is generally located at a larger coordinate than that of the ground state, and thus the lattice usually expands. However, this is not always correct in the case for the RBM coherent phonons of SWCNT, where the tube diameter can either expand or contract depending on the excitation energy. Here we find that such an anomalous behavior occurs as a result of the linear dependence of electronic energy and the details of the electron-phonon interaction.

In order to understand this phenomenon, let us consider the magnitude and phase of the oscillation amplitude $Q(t)$ driven by $S(t)$ in Eq. (5.2). Since $\delta f^\mu \propto M_{op}^\mu$, the magnitude of oscillations should be proportional to the product of the electron-phonon and electron-photon matrix elements:

$$|Q_m| \propto |M_{el\text{-}ph}||M_{OP}|, \qquad (5.4)$$

Since $\delta f^\mu(k)$ is positive if there is no gain in the system, the initial phase of $Q(t)$ is determined only by the sign of $M_{el\text{-}ph}^\mu(k)$ which is summed over all cutting lines $\mu$ and all $k$ points. The values of $|M_{el\text{-}ph}|$ and $|M_{OP}|$ for a fixed selection of energy and $(n,m)$ thus determine the excitation energy and chirality dependence of the coherent phonon amplitudes.

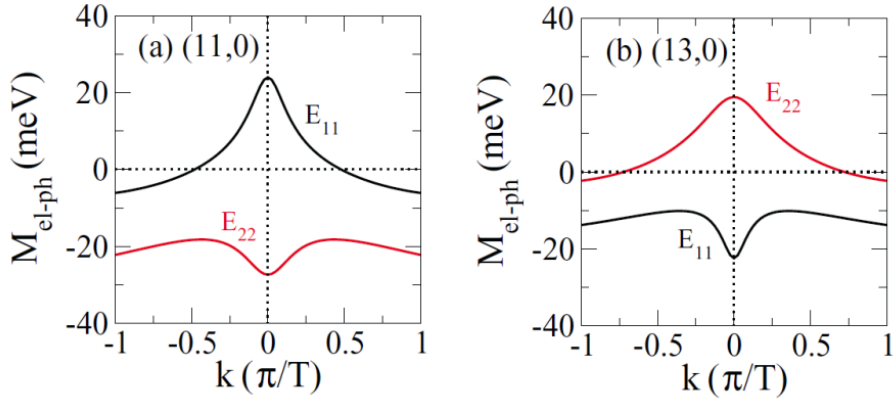

**Figure 32:** RBM electron-phonon matrix elements of (a) (11,0) and (b) (13,0) zigzag nanotubes within the ETB approximation.

We can discuss the type dependence of coherent RBM phonon amplitudes by comparing two semiconducting zigzag nanotubes of different $2n + m$ families and type-I and -II for $\mod(2n + m, 3) = 1$ and $\mod(2n + m, 3) = 2$, respectively. In some references, researchers also use the n - m family notation, in which mod 0, 1, 2 denote $\mod(n - m, 3) = 0, 1, 2$, respectively. In this case, type-I (II) corresponds to mod 2 (1). In Fig. 32, we plot the electron-phonon matrix elements for RBM coherent phonons in the (11,0) (type-I) and (13,0) (type-II) nanotubes as a function of the 1D wave vector $k$. The $k$-dependence of $M^{\mu}_{\text{el-ph}}(k)$ for the RBM phonon is shown for the two cutting lines near the K (or Dirac) point in the hexagonal 2D Brillouin zone, that is, for $E_{11}$ and $E_{22}$. As can be seen in Fig. 32, $M^{\mu}_{\text{el-ph}}(k)$ can be either positive or negative depending on $E_{ii}$ and the nanotubes type. Also, according to Eq. (5.2), if we pump near the $E_{ii}$ band edge, the electron distributions would be localized near $k = 0$ in the 1D Brillouin zone of the zigzag nanotubes, for which the $k_{ii}$ points for the $E_{ii}$ energies lie at $k = 0$. Therefore, the positive (negative) values of $S(t)$ $S(t)$ at the $E_{22}(E_{11})$ transition energy are determined by the negative (positive) values of $M^{\mu}_{\text{el-ph}}(k)$ near $k = 0$. For the two nanotubes, the signs of the electron-phonon matrix elements differ at $E_{11}$ and $E_{22}$. The reason is that for type-I and type-II nanotubes the $E_{11}$ and $E_{22}$ cutting line positions with respect to the K point in the 2D graphene Brillouin zone are opposite to each other [82]. Depending on the cutting line positions relative to the K point, the corresponding $M^{\mu}_{\text{el-ph}}(k)$ for a given cutting line is negative in the region to the right of the K point and positive in the region to the left [83]. This will be proved in the following paragraphs using an effective-mass theory [84]. From this argument, we predict that the type-I (type-II) zigzag nanotubes will start their coherent RBM phonon oscillations by initially decreasing (increasing) the tube diameter at $E_{11}$, while at $E_{22}$ the behavior is just the opposite, as shown in Fig. 32.

The electron-phonon matrix element $M_{\text{el-ph}}$ for the photoexcited electron is basically a sum of conduction band (c) and valence band (v) electron-phonon matrix elements, which represent the electron and hole contributions, respectively [80, 83, 85]:

$$M_{\text{el-ph}} = M^{\text{C}}_{\text{el-ph}} - M^{\text{V}}_{\text{el-ph}} = \langle c|H_{\text{el-ph}}|c\rangle - \langle v|H_{\text{el-ph}}|v\rangle \quad (5.5)$$

where $H_{\text{el-ph}}$ is the SWCNT electron-phonon interaction Hamiltonian. In a nearest-neighbor effective-mass approximation, the RBM $H_{\text{el-ph}}$ for an (n,m) SWCNT with a chiral angle $\theta$ and diameter $d_t$ can be written as [84]:

$$H_{\text{el-ph}} = \frac{2s_r}{d_t} \begin{pmatrix} g_{\text{on}} & -\frac{g_{\text{off}}}{2}e^{i3\theta} \\ -\frac{g_{\text{off}}}{2}e^{-i3\theta} & g_{\text{on}} \end{pmatrix}, \quad (5.6)$$

where $g_{\text{on}}$ ($g_{\text{off}}$) is the on-site (off-site) coupling constant. Here $s_r = \sqrt{\hbar/2M\omega_{\text{RBM}}}$ is the zero point phonon amplitude for the RBM, where $\omega_{\text{RBM}}$ is the phonon frequency and $M$ is the total mass of the carbon atoms within the unit cell. To obtain $M_{\text{el-ph}}$ in Eq. (5.5), we adopt the following two wave functions:

$$\Psi_c = \frac{e^{ikr}}{\sqrt{2S}}\begin{pmatrix} e^{-i\Theta(k)/2} \\ e^{+i\Theta(k)/2} \end{pmatrix}, \quad \Psi_v = \frac{e^{ikr}}{\sqrt{2S}}\begin{pmatrix} e^{-i\Theta(k)/2} \\ -e^{+i\Theta(k)/2} \end{pmatrix}, \quad (5.7)$$

for conduction and valence states, respectively, which are suitable near the graphene K point. In Eq. (5.7), $S$ is the surface area of graphene and $\Theta(\boldsymbol{k})$ is an angle at the K point measured from a line perpendicular to the cutting lines (see Fig. 34). By inserting the wave functions in Eq. (5.7) and the Hamiltonian in Eq. (5.6) into Eq. (5.5), we obtain

$$M_{\text{el-ph}} = \frac{s_r}{d_t}\{-2g_{\text{off}}\cos[\Theta(\boldsymbol{k})] + 3\theta\}. \quad (5.8)$$

Using the results of the density-functional calculation by Porezag et al. [86], we adopt the off-site coupling constant $g_{off} = 6.4$ eV and the on-site coupling constant $g_{on} = 17.0$ eV, which are calculated for the first-nearest-neighbor carbon-carbon distance. However, $g_{on}$ has no effect on the electron-phonon matrix element since it vanishes in Eq. (5.8) if we simply adopt a simple tight binding model in which the valence and conduction energy bands are symmetric with respect to the Fermi energy. If we consider the asymmetry between the valence and conduction bands, the effect of finite $g_{on}$ will appear [87] and give a small correction to Eq. (5.8). Within the present approximation, we do not consider such an asymmetry since the chirality dependence of the electron-phonon matrix element is already described by the $\cos\Theta(k)$ term, which will give a positive or negative sign in front of $g_{off}$ depending on the position of $k$ defined in Fig. 33.

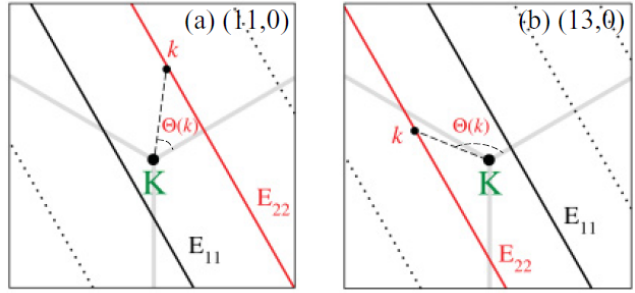

**Figure 33:** Cutting lines for (a) (11,0) and (b) (13,0) zigzag nanotubes near the graphene K point. Black and red solid lines denote the $E_{11}$ and $E_{22}$ cutting lines, respectively, while the dotted lines correspond to higher cutting lines. The angle $\Theta(k)$ is measured counterclockwise from a line perpendicular to the cutting lines, with the positive direction of the line to the right of the K point. Here $\Theta(k)$ is shown for a $k$ point on the $E_{22}$ cutting line for both SWCNTs. The difference between the type-I and type-II families can be understood from the position of the $E_{11}$ or $E_{22}$ cutting lines relative to the K point.

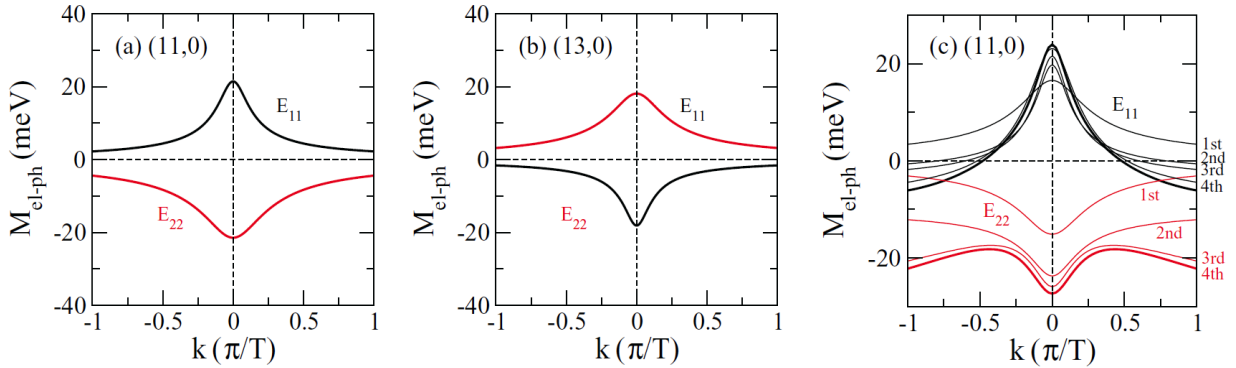

**Figure 34:** RBM electron-phonon matrix elements of (a) (11,0) and (b) (13,0) nanotubes calculated within the effective-mass theory using $g_{off} = 6.4$ eV. In (a) and (b), the matrix elements near $k = 0$ are comparable with the results in Figure 5.2. Panel (c) shows the matrix elements of an (11,0) nanotube calculated within the ETB model for interactions up to the fourth-nearest neighbors. The results including fourth-nearest neighbors exactly reproduce the results in Figure 5.2(a).

For zigzag nanotubes, Eq. (5.8) also explains the dependence of $M_{el-ph}$ on the cutting line (or $k$) position. Let us take the examples in Fig. 33, in which we show the cutting lines for the (11,0) and (13,0) nanotubes. The $E_{22}$ cutting line for the (11,0) [(13,0)] tube is to the right (left) of the K point, giving a positive (negative) $\cos\Theta(k)$ and thus a negative (positive) $M_{el-ph}$ for the $E_{22}$ transition. According to Eq.

(5.2), the negative (positive) $M_{\text{el-ph}}$ corresponds to an initial increase (decrease) of the tube diameter. In this way, the chirality dependence of the coherent phonon amplitude is simply determined by the electron-phonon interaction. However, we should note that this simple rule does not work well for $E_{33}$ and $E_{33}$, as can be seen in Fig. 31. For instance, the coherent phonon amplitude at $E_{33}$ has the same sign as that at $E_{44}$ although their cutting line positions are opposite each other with respect to the K point. The reason for the breakdown of this simple rule is that the cutting lines for $E_{33}$ and $E_{44}$ are far from the K point so that the wave functions of Eq. (5.7) are no longer good approximations. In this case, the ETB wave functions (*not* the effective-mass ones) are necessary for obtaining the coherent phonon amplitudes.

In Fig. 34, we then plot the electron-phonon matrix elements of Eq. (5.8) for the (11,0) and (13,0) nanotubes, where the on-site term ($g_{\text{on}}$) disappears and only the off-site term ($g_{\text{off}}$) contributes to $M_{\text{el-ph}}$. It can be seen that the effective-mass theory [see Figs. 34(a) and 34(b)] reproduces the ETB calculation results only near $k_{ii} = 0$ [Figs. 32(a) and 32(b)]. However, the first-nearest-neighbor effective-mass model [Figs. 34(a) and 34(b)] cannot reproduce the ETB matrix element results (Fig. 32) at $k$ far from $k_{ii} = 0$ because in the ETB calculation we consider not only asymmetry of the energy bands but also electron-phonon interaction up to fourth nearest neighbors. To stress this fact, in Fig. 33(c) we show the $k$ dependence of $M_{\text{el-ph}}$ for the (11,0) tube within the ETB model for interactions up to the fourth-nearest neighbors. Based on this figure, we conclude that a more exact analytical expression for $M_{\text{el-ph}}$ at $k$ far from the $k_{ii}$ should take into account the longer-range electron-phonon interactions. Nevertheless, the first-nearest-neighbor effective-mass theory has already given physical insight into the $k$-dependent $M_{\text{el-ph}}$.

To consider the more general family behavior of the RBM coherent phonon amplitudes, we recalculated $Q_m$ using the ETB program for 31 different SWCNT chiralities with diameters of 0.7–1.1 nm and for photo excitations at $E_{ii}$ in the range 1.5–3.0 eV. The results are shown in Fig. 35. Note that, in addition to the semiconducting SWCNTs, we also give some results for metallic SWCNTs. It is known that the densities of states for $E_{ii}$ in metallic SWCNTs are split into lower $E_{ii}^{\text{L}}$ and higher $E_{ii}^{\text{H}}$ branches, except for the armchair SWCNTs [88]. Here we consider $Q_m$ in metallic SWCNTs only at $E_{11}^{\text{L}}$. The cutting line for $E_{11}^{\text{L}}$ is located to the right of the K point. We can see in Fig. 35 that all the metallic SWCNTs start vibrations by increasing their diameters at $E_{11}^{\text{L}}$. The reason is the same as in type-II nanotubes, where the cutting lines for the $E_{11}$ transitions are located to the right of the K point, giving a negative $M_{\text{el-ph}}$ (hence a positive $Q_m$) as explained within the effective-mass theory. On the other hand, at $E_{11}^{\text{H}}$, the nanotubes should start their coherent vibrations by decreasing their diameters. In the case of armchair nanotubes, for which $E_{11}^{\text{L}} = E_{11}^{\text{H}}$, we expect that no vibration should occur because the two contributions from $E_{11}^{\text{L}}$ and $E_{11}^{\text{H}}$ should cancel each other.

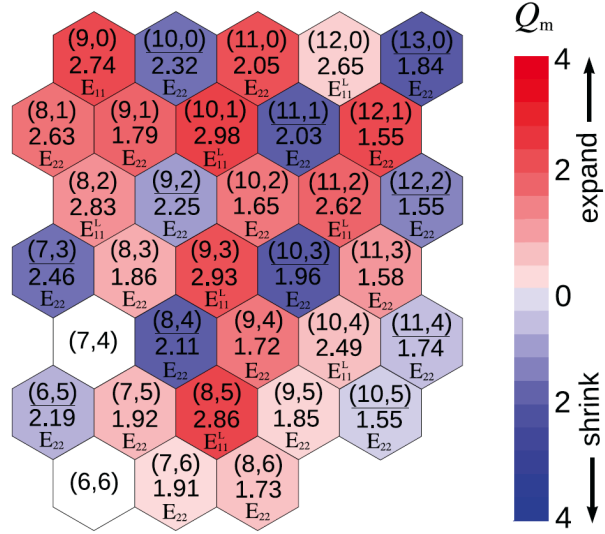

**Figure 35:** The lattice response of SWCNTs with diameters in the range 0.7–1.1 nm is mapped onto the unrolled graphene lattice specifiying the tube chiralities (*n,m*). In this map $Q_m$ is expressed in terms of $\sqrt{\hbar/2M\omega_{RBM}}$. Red and blue colored hexagons denote the SWCNTs whose vibrations start by increasing or decreasing their diameters, respectively. For clarity, the shrinking tubes (blue colored hexagons) are also specified by underlining their (*n,m*). The laser excitation energies are selected within the range 1.5–3.0 eV. For each (*n,m*) tube, the corresponding $E_{ii}$ (in eV) found within this energy region is listed on each hexagon with the label $E_{ii}$. The calculated results for the (7,4) and (6,6) nanotubes are not shown in this figure because their $E_{11}^L > 3.0$ eV and the (6,6) tube gives a negligibly small $Q_m$.

For semiconducting nanotubes, we see that most of the type-I (type-II) nanotubes start vibrating at $E_{11}$ by decreasing (increasing) their diameters and at higher energies by increasing (decreasing) their diameters. In a few cases, e.g., (7,6), (9,5), and (10,5) nanotubes, the deviation from this rule might come from the $3\theta$ term in Eq. (5.8), especially for the near-armchair nanotubes where $\theta$ approaches $\pi/6$. As mentioned previously, we consider that in the case of armchair nanotubes, for example the (6,6) nanotube, which is metallic, the coherent phonon amplitude becomes small because of the trigonal warping effect [88]. The exclusion of both excitonic and environmental effects may also be a reason for this deviation because the $E_{ii}$ transition energies are also shifted to some extent. Nevertheless, our results should stimulate further work by experimentalists to check for consistency with this prediction.

*5.2 Coherent RBLM phonons in high symmetry GNRs.*

Next, we extend our coherent phonon theory to the cases of zigzag graphene nanoribbons (zGNRs) and armchair graphene nanoribbons (aGNRs). In GNRs, the rotational degree of freedom about the nanotube axis is lost and the number of CP active phonon modes is equal to the number of AB carbon dimers in the nanoribbon translational unit cell. The most easily observed of these CP modes is the one with the lowest frequency, namely the radial breathing like mode (RBLM) in which the width of the GNR is vibrating.

The lattice structure for aGNRs and zGNRs is shown schematically in Fig. 36. These ribbons are denoted $N_{ab}$ aGNR and $N_{ab}$ zGNR, respectively, where $N_{ab}$ is the number of AB carbon dimers in the translational unit cell [89, 90]. In zigzag ribbons, the length $L$ of the translational unit cell is $a$ and the width $W$ of the ribbon is $(N_{ab}-1)\frac{\sqrt{3}}{2}a$, where $a = 2.49$ Å is the hexagonal lattice constant in graphene. In armchair ribbons, the translational unit cell length is $\sqrt{3}a$ and the ribbon width is $(N_{ab}-1)\frac{1}{2}a$. Note that in zigzag and armchair ribbons with the same number of atoms per unit cell, the area of the unit cells are equal.

*5.2.1 Zigzag GNRs*

First we consider zGNRs. Using a treatment similar to that in the nanotube case, we discuss the initial expansion and contraction of the nanoribbon width in coherent radial breathing like mode (RBLM) oscillations. These nanoribbons are basically metallic because of the edge states at the Fermi energy in the zGNRs [91].

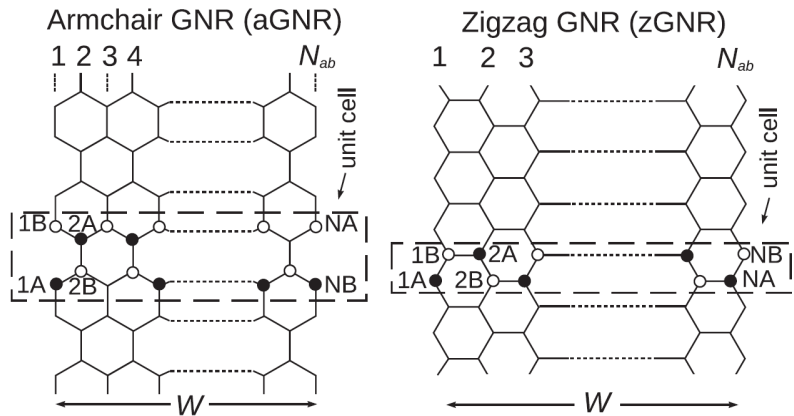

**Figure 36:** Schematic diagram showing lattice structures and translational unit cells for armchair (aGNR) and zigzag (zGNR) graphene nanoribbons. The width of the nanoribbons is $W$.

Since coherent phonon spectroscopy gives direct phase information on the coherent phonon amplitude, it is instructive to examine $S_{max}$, i.e., the maximum value of $S(t)$ as a function of pump photon energy. Figure 37(a) is the power spectrum of $Q(t)$ at the RBLM frequency. In Fig. 37(b) we plot $S_{max}$ as a function of pump photon energy. For comparison, the absorption coefficient is plotted in Fig. 37(c). Near the band edge, we see from Fig. 37(b) that the pump light is strongly absorbed at the V2C1 and V3V1 peaks. Here V$n$C$m$ denotes transitions between the low lying $n^{th}$ hole band and the higher lying $m^{th}$ electron band. The resulting increase in the photoexcited carrier density increases the coherent phonon driving function and enhances the coherent phonon oscillation amplitudes. In other words, the coherent phonon driving function near the band edge is determined by the strength of optical absorption between the lowest few hole bands and the localized edge states V1 and C1. At energies above 3 eV, $S_{max}$ changes sign and the nanoribbon initially contracts.

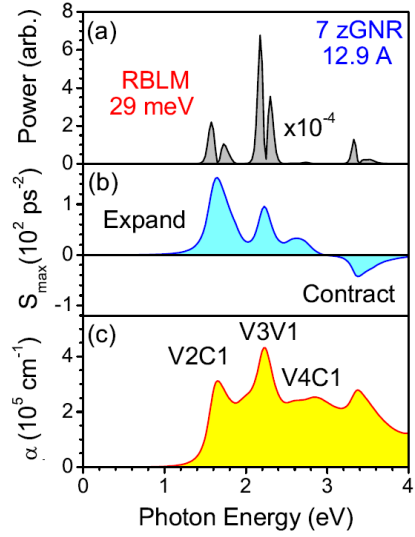

**Figure 37:** (a) Coherent phonon power at the RBLM frequency (29 meV), (b) The value of $S_{max}$, and (c) Initial absorption spectrum as a function of photon energy for 7 zGNR.

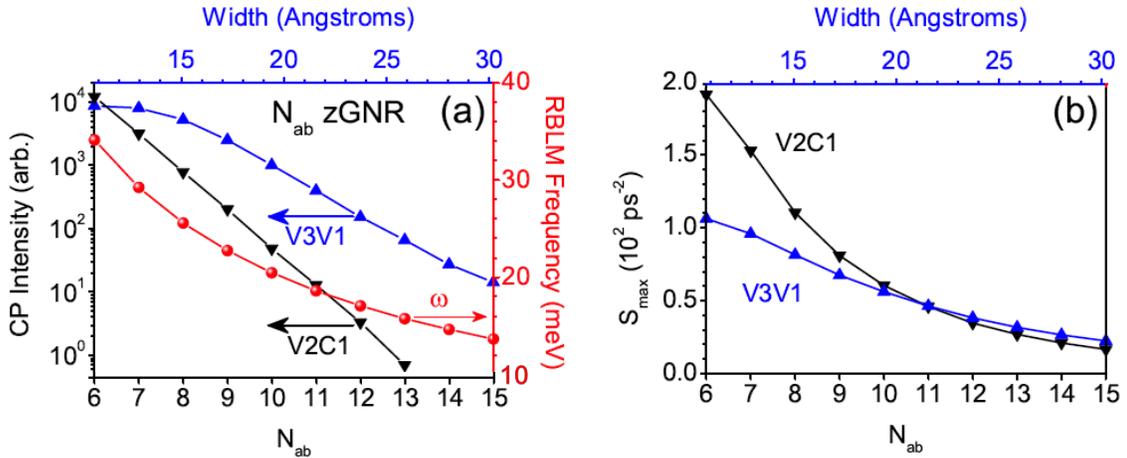

**Figure 38:** For zigzag nanoribbons excited by Gaussian laser pulse with pump and probe polarization vector parallel to ribbon length, in (a) we plot the RBLM CP intensity for the V2C1 and V3V1 transitions as a function of $N_{ab}$, the number of carbon dimers in the zigzag unit cell, on the left axis. On the right axis in we plot the RBLM frequency $\omega$ in eV. In (b) we plot $S_{max}$ for the V2C1 and V3V1 transitions. The ribbon width for each value of $N_{ab}$ can be read from the upper axis.

It is important to plot the CP intensity as a function of nanoribbon width because the RBLM CP amplitude of zGNRs quickly decreases with increasing nanoribbon width. We fix the pump and probe electric polarization vectors to be parallel to the nanoribbon axis as this is the polarization for which the CP intensity is greatest. In Fig. 38(a), the RBLM CP intensity for the V2C1 and V3V1 transitions as a function of the number of carbon dimers in the zigzag nanoribbon unit cell is plotted against the left axis and the RBLM frequency is plotted against the right axis. In Fig. 38(b), we plot the coherent driving function amplitude $S_{max}$ as a function of $N_{ab}$ for the V2C1 and C3V1 transitions. We have studied all zGNR nanoribbons for $N_{ab}$ ranging from 6 to 17 and find similar results. The driving function is positive for low energies (up to just below 3 eV) and then becomes negative for all cases.

*5.2.2 Armchair GNRs*

Armchair GNRs belong to one of three families depending on the mod number: mod($N_{ab}$, 3). Based on a simple band structure calculation, we classify mod 0 and mod 1 aGNRs as semiconductors and mod 2 aGNRs as metals [89]. However, unlike the case in zigzag nanoribbons [91], there are no localized edge states near the band edge. Armchair nanoribbons have direct gaps that arise from quantum confinement and edge effects and all the electronic wave functions near the band edge are distributed throughout the width of the ribbon.

We examine $S_{max}$ as a function of pump photon energy and our results for a 6 aGNR mod 0 semiconducting nanoribbon are shown in Fig. 39(a), where $S_{max}$ is shown as a function of pump photon energy. For comparison, the absorption coefficient is also plotted in the lower panel of Fig. 39(a). Near the band edge, we see from Fig. 39(a) that the pump light is strongly absorbed near the $E_{11}$ and $E_{22}$ peaks. The resulting increase in the photoexcited carrier density increases the coherent phonon driving function and enhances the coherent phonon oscillation amplitudes. Photoexcitation by the pump causes the nanoribbon to initially expand for pump photon energies near the $E_{11}$ transition and to initially contract for pump photon energies near the $E_{22}$ transition. We find this to be true for all mod 0 semiconducting nanoribbons. Qualitatively, different results are obtained for mod 1 aGNRs. In Fig. 39(b) we plot $S_{max}$ as a function of pump photon energy for a 7 aGNR mod 1 nanoribbon and find that photoexcitation by the pump causes the nanoribbon to initially contract for photon energies near the $E_{11}$ peak and initially expand for photon energies near the $E_{22}$ peak. This is found to be true for all mod 1 semiconducting nanoribbons. In Fig. 39(c) we show results for an 8 aGNR mod 2 metallic nanoribbon excited by a laser pulse polarized parallel to the ribbon length. From Fig. 39(c), we see that photoexcitation by the pump causes the nanoribbon to initially expand for photon energies near the $E_{11}$ transition. For photon energies near the $E_{22}$ transition, the situation is more ambiguous.

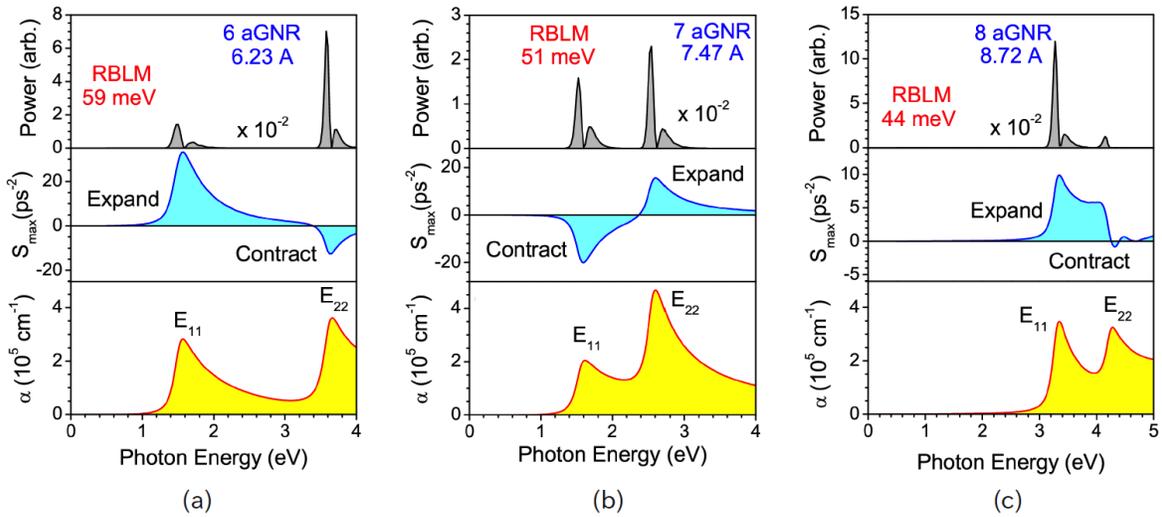

**Figure 39:** The coherent phonon power, the value of Smax, and the initial absorption spectrum are plotted as a function of photon energy for (a) 6 aGNR mod 0 nanoribbon (RBLM frequency = 59 meV), (b) 7 aGNR mod 1 nanoribbon (RBLM frequency = 51 meV), and (c) 8 aGNR mod 2 nanoribbon (RBLM frequency = 44 meV). The excitation is due to a Gaussian laser pulse with pump and probe polarization vectors parallel to the ribbon length.

Following our previous study on the nanotube system [9], we can also examine the *k*-dependent electron-phonon interaction within the effective mass approximation to explain why some GNRs start their coherent RBLM oscillations by initially expanding while others start the oscillations by initially shrinking. In the present discussion we will focus our attention on the aGNRs, in which we can directly use the wavefunctions formulated in Eq. (5.7). In the case of zGNRs, we have to consider a special localized wavefunction due to the presence of edge states at which the $E_{ii}$ transition occurs. Using such a wavefunction, we obtain a constant electron-phonon matrix element that does not depend on mod($N_{ab}$, 3) of the zGNRs, and thus is consistent with our results in Section 5.2.1.

By a similar calculation as in Eqs. (5.6)-(5.8), we obtain the electron-phonon matrix element for the RBLM in aGNRs:

$$M_{\text{el-ph}} = u_{\text{arm}}[-2g_{\text{off}}\cos\Theta(\mathbf{k})] \quad (5.8)$$

where $u_{\text{arm}}$ is a ribbon width- or $N_{ab}$-dependent phonon amplitude and $\Theta(\mathbf{k})$ as defined in Fig. 40 (see also Appendix D in Ref. [10]). From Eq. (5.8), we can analyze the $N_{ab}$ and $E_{ii}$ dependence of the aGNR initial lattice response. First of all, we should note that $g_{\text{off}} = 6.4$ eV and $u_{\text{arm}}$ are always positive [10], while $\cos\Theta(\mathbf{k})$ can either be positive or negative depending on the value of $k$ at which the $E_{ii}$ transition occurs. According to the definition of the driving force in Eqs. (5.1) and (5.2), a negative (positive) $M_{\text{el-ph}}$ value corresponds to a positive (negative) $S_{\text{max}}$. Therefore, a positive (negative) $\cos\Theta(\mathbf{k})$ is related to a contraction (expansion) of the ribbon width. Using this argument, we can classify the aGNR lattice response based on the aGNR types. For example, let us consider semiconducting mod 0 aGNRs and mod 1 aGNRs. The cutting line position for their $E_{11}$ and $E_{22}$ optical transitions are just opposite to each other. For a mod 0 aGNR, we see that $\cos\Theta(\mathbf{k})$ becomes positive (negative) at $E_{11}(E_{22})$, and thus the aGNR starts the coherent phonon oscillations by expanding (shrinking) its width. This can be seen in the illustration of $\Theta(\mathbf{k})$ in Fig. 41. The opposite behavior is true for mod 1 aGNRs.

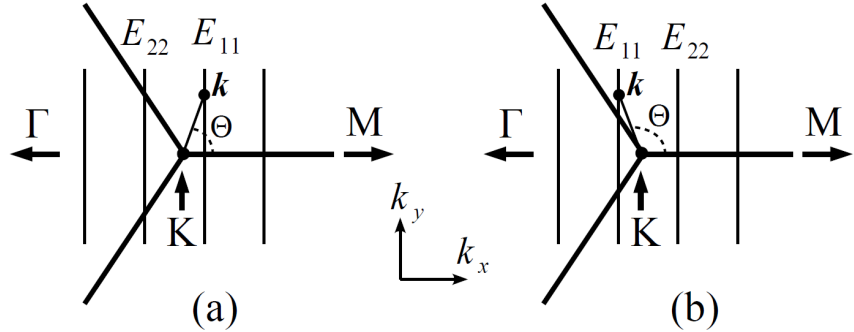

**Figure 40:** Cutting lines for (a) mod 0 aGNR and (b) mod 1 aGNRs near the Dirac K point. To make clear the definition of $\Theta(\mathbf{k})$, in this figure is shown for an arbitrary $k$ at $E_{11}$. In fact, in the case of mod 0 and mod 1 aGNRs the $E_{11}$ transitions occur at $\Theta(\mathbf{k}) = 0$ and $\Theta(\mathbf{k}) = \pi$, respectively. The difference between the mod 0 and mod 1 aGNRs can be understood from the position of the $E_{11}$ or $E_{22}$ cutting lines relative to the K point.

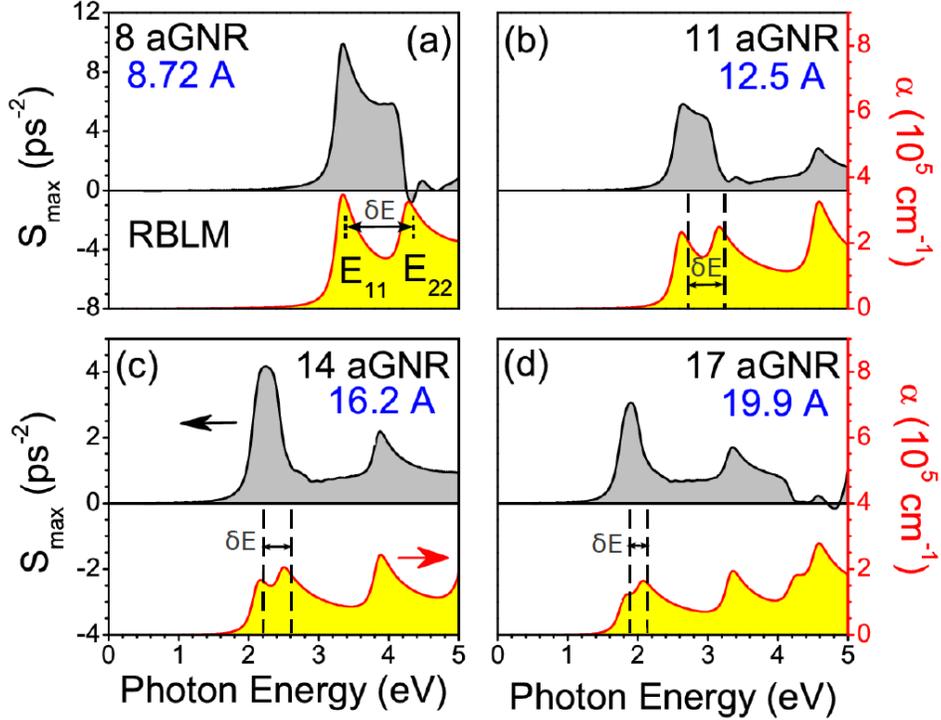

**Figure 41:** Driving force $S_{max}$ and initial absorption spectrum plotted as a function of photon energy for several mod 2 metallic aGNRs: (a) $N_{ab}= 8$, (b) $N_{ab}= 11$, (c) $N_{ab}= 14$, and (d) $N_{ab}= 17$. Positive (negative) $S_{max}$ at $E_{11}(E_{22})$ corresponds to an expansion (contraction) of the ribbon width.

However, the driving force trends for the mod 2 metallic aGNRs (see Fig. 41) cannot be explained nicely by the effective mass theory for several reasons. The main reason is that, in the metallic aGNRs, there are two cutting lines with the same distance from the K point, which can be assigned as the lower and higher branches of an $E_{ii}$ transition. Both branches contribute to a specific $E_{ii}$ and thus we have to sum up the matrix elements from each contribution to obtain $M_{\text{el-ph}}$. For example, if the 1D k-points for the lower and higher branches of $E_{ii}$ are the same, the matrix elements will cancel each other because $\cos\Theta(\mathbf{k})+\cos(\pi-\Theta(\mathbf{k}))=0$. In this case, the CP amplitude will be generally small for the mod 2 metallic aGNRs compared to the mod 0 or mod 1 semiconducting aGNRs. In the realistic situation, we always have slightly different k-points for the lower and higher branches due to the trigonal warping effect [88], from which the nonzero $M_{\text{el-ph}}$ value gives information about an expansion or contraction of the ribbon width.

Near the $E_{11}$ transition, the metallic aGNR initial lattice response is always an expansion for all $N_{ab}$. On the other hand, near the $E_{22}$ transition, the response is expected to be always a contraction, though we see in Figs. 41(b)-(d) the trend does not hold for larger $N_{ab}$. We notice that the difference $\delta E = E_{22} - E_{11}$ might determine whether or not the lattice response at the $E_{22}$ feature of a given mod 2 $N_{ab}$ aGNRs will clearly follow our effective mass theory. We guess that if $\delta E$ is large enough (around 2 eV as in the 8 aGNR), the lattice response at $E_{22}$ should not be ambiguous. However, $\delta E$ decreases with increasing $N_{ab}$ as can be understood from a cutting line argument [82]. Thus, the lattice response at $E_{22}$ for the larger mod 2 aGNRs becomes opposite to that for the smaller mod 2 aGNRs (e.g., $N_{ab} = 2, 5, 8$).

We should also note that the $E_{11}$ and $E_{22}$ values of the mod 2 metallic aGNRs are close to the order of $E_{33}$ and $E_{44}$ values for the mod 0 and mod 1 semiconducting aGNRs. This gives another reason why the effective mass theory cannot explain the metallic aGNR trends. In this energy region, the nearest-neighbor effective mass theory should be extended to include longer-range nearest-neighbor interactions.

Finally, we summarize the lattice behavior at $E_{11}$ and $E_{22}$ transitions for all families of aGNRs in the following table:

| family | $E_{11}$ | $E_{22}$ |
|---|---|---|
| mod 0 | expand | contract |
| mod 1 | contract | expand |
| mod 2 | expand | expand or contract |

## 6. Summary

We have generated and detected coherent phonon lattice vibrations in SWCNTs and graphene corresponding to the RBM and G-mode using ultrashort laser pulses. Coherent phonon spectroscopy has several advantages, including excellent resolution and narrow line widths, no Rayleigh scattering and no photoluminescence signal, and direct measurement of the vibrational dynamics, which contains phase information and decay times. The ability of CP measurements to trace the first derivative of the excitonic absorption peaks of specific chirality (*n,m*) tubes will allow in-depth study of the line shape of these resonances. In addition, chirality-selective excitations of coherent RBM oscillations in SWCNTs can be produced by implementing multiple pulse trains with repetition rates matched to (*n,m*) specific RBM frequencies. We obtained single-chirality information for many distinct tube chiralities and extracted detailed information about the phase and modulation of the absorption, leading to an experimental confirmation that for (11,3) SWCNT's the lattice initially expands in response to the pump pulse.

We have demonstrated that coherent G-mode phonons having a resonant frequency of approximately 47.7 THz are generated by an impulsively stimulated Raman scattering (ISRS) process. Two different detection mechanisms in single-walled carbon nanotubes (SWCNTs) and graphene were used. In SWCNTs, coherent G-mode phonon oscillations were detected through stimulated Stokes and anti-Stokes Raman scattering process by performing spectrally-resolved pump-probe measurements. The probe energy dependence of the oscillation amplitude and the preferential occurrence between SSRS and SARS for the chirped pulses were explained within the mechanism of ISRS with spectrally broad pulses.

When a pulse to pulse interval in multiple pulse trains corresponds to the period of a specific Raman mode frequency, the frequency that are matched to the repetition rate of the multiple pulse trains was generated with increasing oscillatory response to each pulse, while others were diminished coherent response. Using the tailored optical pulse trains generated by a pulse-shaping technique, we selectively excited and probed coherent lattice vibrations of the radial breathing modes at around 6-8 THz of specific chirality SWCNTs within an ensemble sample of various different chiralities. Such coherent phonon signals provide chirality-specific information about the phase and modulation of the absorption as a function of time.

To explain our experimental results, we have developed a microscopic theory using a tight-binding model for the electronic states and a valence force field model for the phonons. We find that the CP amplitudes satisfy a driven oscillator equation with the driving term depending on photoexcited carrier density. We find that the RBM CP amplitudes are very sensitive to changes in excitation energy and depend strongly on chirality. Our model predicts the overall trends in the relative strengths of the CP signal both within and between different (2*n*+*m*) families. We also theoretically analyze how the tube diameter changes in response to femtosecond laser excitation and under what conditions the diameter will initially increase or decrease. Finally, we developed a microscopic theory for CP generation and detection in graphene nanoribbons. We examine the CP radial breathing like mode (RBLM) amplitudes as a function of excitation energies and nanoribbon type. For photoexcitation near the optical absorption edge, the CP driving term for the RBLM is much larger for zigzag nanoribbons where the strong transitions between localized edge states provide the dominant contribution to the CP driving term. Using an effective mass theory, we explain how the armchair nanoribbon width changes in response to laser excitation.

**Acknowledgements**

This work was supported by the National Science Foundation under Grants No. DMR-1105437 and No. OISE-0968405, the Robert A. Welch Foundation through Grant No. C-1509, and the Office of Naval Research under Grant No. ONR- 00075094. R.S. acknowledges a MEXT grant (Ministry of Education, Japan, No. 20241023). Y.L. and K.Y. acknowledge the support by the NRF grant (2010-0022691, 2010-0028165).

**References**

[1] Y.-S. Lim, K.-J. Yee, J.-H. Kim, E.H. Hároz, J. Shaver, J. Kono, S.K. Doorn, R.H. Hauge, R.E. Smalley, Nano Lett., 6 (2006) 26-96.
[2] J.-H. Kim, K.-J. Han, N.-J. Kim, K.-J. Yee, Y.-S. Lim, G.D. Sanders, C.J. Stanton, L.G. Booshehri, E.H. Hároz, J. Kono, Phys. Rev. Lett., 102 (2009) 037402.
[3] J.-H. Kim, J.-G. Park, B.-Y. Lee, D.-H. Lee, K.-J. Yee, Y.-S. Lim, L.G. Booshehri, J.K. E. H. Hároz, and S.-H. Baik, J. Appl. Phys., 105 (2009) 103506.
[4] Y.-S. Lim, J.-G. Ahn, J.-H. Kim, K.-J. Yee, T. Joo, S.-H. Baik, E.H. Haroz, L.G. Booshehri, J. Kono, ACS Nano, 4 (2010) 3222.
[5] L.G. Booshehri, C.L. Pint, G.D. Sanders, L. Ren, C. Sun, E.H. Hároz, J.-H. Kim, K.-J. Yee, Y.-S. Lim, R.H. Hauge, C.J. Stanton, J. Kono, Phys. Rev. B, 83 (2011) 195411.
[6] J.-H. Kim, K.-J. Yee, Y.-S. Lim, L.G. Booshehri, E.H. Hároz, J. Kono, arXiv:1106.0838v1.
[7] G.D. Sanders, C.J. Stanton, J.-H. Kim, K.-J. Yee, Y.-S. Lim, E.H. Hároz, L.G. Booshehri, J. Kono, R. Saito, Phys. Rev. B, 79 (2009) 205-434.
[8] A.R.T. Nugraha, R. Saito, K. Sato, P.T. Araujo, A. Jorio, M.S. Dresselhaus, Appl. Phys. Lett., 97 (2010) 091905.
[9] A. R. T. Nugraha, G. D. Sanders, K. Sato, C. J. Stanton, M.S. Dresselhaus, R. Saito, Phys. Rev. B, 84 (2011) 174302.
[10] G.D. Sanders, A.R.T. Nugraha, R. Saito, C.J. Stanton, Phys. Rev. B, 85 (2012) 205401.
[11] W.A. Kutt, W. Albrecht, H. Kurz, IEEE J. Quant. Electron., 28 (1992) 2434.
[12] R. Merlin, Solid State Comm., 102 (1997) 207.
[13] A.V. Kuznetsov, C.J. Stanton, Theory of Coherent Phonon Oscillations in Bulk GaAs in Ultrafast Phenomena in Semiconducotrs, Springer-Verlag, New York, 2001.
[14] A.V. Kuznetsov, C.J. Stanton, Phys. Rev. B, 51 (1995) 7555.
[15] H.J. Zeiger, J. Vidal, T.K. Cheng, E.P. Ippen, G. Dresselhaus, M.S. Dresselhaus, Phys. Rev. B, 45 (1992) 768.
[16] G.C. Cho, W. Kutt, H. Kurz, Phys. Rev. Lett., 65 (1990) 764-766.
[17] W. Kutt, G.C. Cho, T. Pfeifer, H. Kurz, Semicond. Sci. Technol., 7 (1992) B77.
[18] T. Pfeifer, W. Kutt, H. Kurz, H. Scholz, Phys. Rev. Lett., 69 (1992) 3248.
[19] T.K. Cheng, J. Vidal, H.J. Zeiger, G. Dresselhaus, M.S. Dresselhaus, E.P. Ippen, Appl. Phys. Lett., 59 (1991) 1923.
[20] W. Albrecht, T. Kruse, H. Kurz, Phys. Rev. Lett., 69 (1992) 1451.
[21] F.C.G. Sudarshan, J.R. Klauder, Fundamentals of Quantum Optics, Benjamin, New York, 1968.
[22] C. Cohen-Tannoudji, B. Diu, F. Laloe, Quantum Mechanics, Wiley, New York, 1977.
[23] A. Lauberau, W. Kaiser, Rev. Mod. Phys., 50 (1978) 607.
[24] K. Kato, K. Ishioka, M. Kitajima, J. Tang, R. Saito, H. Petek, Nano Lett., 8 (2008) 3102-3108.
[25] S. Koga, I. Katayama, S. Abe, H. Fukidome, M. Suemitsu, M. Kitajima, J. Takeda, Appl. Phys. Exp., 4 (2011) 045101.
[26] A.V. Kuznetsov, C.J. Stanton, Phys. Rev. Lett., 73 (1994) 3242-3246.
[27] H. Ajiki, T. Ando, Physica B, 201 (1994) 349-352.
[28] V.N. Popov, L. Henrard, Phys. Rev. B 70 (2004) 115407.

[29] D.W. Bailey, C.J. Stanton, Appl. Phys. Lett., 60 (1992) 880.
[30] S. Zollner, K.D. Myers, K.G. Jenson, J.M. Dolan, D.W. Bailey, C.J. Stanton, Solid State Commun., 104 (1997) 51.
[31] S. Byun, H. Jeong, K. Min, S. Byun, C.J. Stanton, D.H. Reitze, J.K. Yoo, G.C. Yi, Y.D. Jho, Appl. Phys. Lett., 100 (2012) 092106.
[32] C.J. Cook, S. Khan, G.D. Sanders, X. Wang, D.H. Reitze, Y.D. Jho, Y.-W. Heo, J.-M. Erie, D.P. Norton, C.J. Stanton, Proc. SPIE, 7603 (2010) 760304.
[33] R. Saito, G. Dresselhaus, M.S. Dresselhaus, Physical Properties of Carbon Nanotubes, Imperial College Press, London, 1998.
[34] K.-A. Wang, A.M. Rao, P.C. Eklund, M.S. Dresselhaus, G. Dresselhaus, Phys. Rev. B, 48 (1993) 11375-11380.
[35] R. Saito, A. Gruneis, G.G. Samsonidze, V.W. Brar, G. Dresselhaus, M.S. Dresselhaus, A. Jorio, L.G. Cancado, C. Fantini, M.A. Pimenta, A.G. Souza Filho, New J. Phys., 5 (2003) 157.
[36] G.G. Samsonidze, E.B. Barros, R. Saito, J. Jiang, G. Dresselhaus, M.S. Dresselhaus, Phys. Rev. B, 75 (2007) 155420.
[37] J. S. Park, K. Sasaki, R. Saito, W. Izumida, M. Kalbac, H. Farhat, G. Dresselhaus, M.S. Dresselhaus, Phys. Rev. B, 80 (2009) 081402.
[38] H. Farhat, K. Sasaki, M. Kalbac, M. Hofmann, R. Saito, M. S. Dresselhaus, J. Kong, Phys. Rev. Lett., 102 (2009) 126804.
[39] M. Lazzeri, C. Attaccalite, L. Wirtz, F. Mauri, Phys. Rev. B, 78 (2008) 081406.
[40] D. Graf, F. Molitor, K. Ensslin, C. Stampfer, A. Jungen, C. Hierold, L. Wirtz, Nano Lett., 7 (2007) 238-242.
[41] W. Kohn, Phys. Rev. Lett., 2 (1959) 393-394.
[42] M.S. Strano, S.K. Doorn, E.H. Haroz, C. Kittrell, R.H. Hauge, R.E. Smalley, Nano Lett., 3 (2003) 1091-1096.
[43] A. Jorio, R. Saito, J. H. Hafner, C. M. Liber, M. Hunter, T. McClure, G. Dresselhaus, M. S. Dresselhaus, Phys. Rev. Lett., 86 (2001) 1118.
[44] M.S. Dresselhaus, G. Dresselhaus, A. Jorio, A.G.S. Filho, R. Saito, Carbon, 40 (2002) 2043-2061.
[45] A. Gambetta, C. Manzoni, E. Menna, M. Meneghetti, G. Cerullo, G. Lanzani, S. Tretiak, A. Piryatinski, A. Saxena, R.L. Martin, A.R. Bishop, Nature Phys., 2 (2006) 515-520.
[46] R.B. Weisman, S.M. Bachilo, Nano Lett., 3 (2003) 1235-1238.
[47] A. Jorio, A.G.S. Filho, G. Dresselhaus, M.S. Dresselhaus, A.K. Swan, M.S. Ünlü, B.B. Goldberg, M.A. Pimenta, J.H. Hafner, C.M. Lieber, R. Saito, Phys. Rev. B, 65 (2002) 155412.
[48] R. Saito, A. Jorio, J. H. Hafner, C. M. Lieber, M. Hunter, T. McClure, G. Dresselhaus, M.S. Dresselhaus, Phys. Rev. B, 64 (2001) 085312.
[49] E.H. Hároz, J.G. Duque, W.D. Rice, C.G. Densmore, J. Kono, S.K. Doorn, Phys. Rev. B, 84 (2011) 121403(R).
[50] K. Sasaki, R. Saito, G. Dresselhaus, M. S. Dresselhaus, H. Farhat, J. Kong, Phys. Rev. B, 78 (2008) 235405.
[51] A. Jorio, M.S. Dresselhaus, R. Saito, G. Dresselhaus, Raman Spectroscopy of Graphene Related Systems, Wiley, 2011.
[52] A.M. Weiner, J.P. Heritage, E.M. Kirschner, J. Opt. Soc. Am. B, 5 (1988) 1563-1572.
[53] A.M. Weiner, D.E. Leaird, Optics Lett., 15 (1990) 51-53.
[54] S. Manfred, Number Theory in Science and Communication, Springer-Verlag, Berlin, 2009.
[55] R. Skaug, R. Skaug, J. Hjelmstad, Spread Spectrum in Communication, Peregrinus, London, 1985.
[56] A.M. Weiner, D.E. Leaird, G.P. Wiederrecht, K.A. Nelson, Science, 247 (1990) 1317-1319.
[57] K.J. Yee, Y.S. Lim, T. Dekorsy, D.S. Kim, Phys. Rev. Lett., 86 (2001) 1630-1633.
[58] M.J. O'Connell, S.M. Bachilo, C.B. Huffman, V.C. Moore, M.S. Strano, E.H. Haroz, K.L. Rialon, P.J. Boul, W.H. Noon, C. Kittrell, J. Ma, R.H. Hauge, R.B. Weisman, R.E. Smalley, Science, 297 (2002) 593-596.

[59] T. Laarmann, I. Shchatsinin, A. Stalmashonak, M. Boyle, N. Zhavoronkov, J. Handt, R. Schmidt, C.P. Schulz, I.V. Hertel1, Phys. Rev. Lett., 98 (2007) 058302.
[60] K.G. Lee, D.S. Kim, K.J. Yee, H.S. Lee, Phys. Rev. B, 74 (2006) 113201.
[61] M. Hase, K. Mizoguchi, H. Harima, S. Nakashima, M. Tani, K. Sakai, M. Hangyo, Appl. Phys. Lett., 69 (1996) 2474.
[62] G.N. Ostojic, S. Zaric, J. Kono, M.S. Strano, V.C. Moore, R.H. Hauge, R.E. Smalley, Phys. Rev. Lett., 92 (2004) 117402.
[63] S.M. Bachilo, M.S. Strano, C. Kittrell, R.H. Hauge, R.E. Smalley, R.B. Weisman, Science, 298 (2002) 2361-2366.
[64] H.H. Gommans, J.W. Alldredge, H. Tashiro, J. Park, J. Magnuson, A.G. Rinzler, J. Appl. Phys., 88 (2000) 2509.
[65] G.S. Duesberg, I. Loa, M. Burghard, K. Syassen, S. Roth, Phys. Rev. Lett., 85 (2000) 5436-5439.
[66] A. Jorio, G. Dresselhaus, M.S. Dresselhaus, M. Souza, M.S.S. Dantas, M.A. Pimenta, A.M. Rao, R. Saito, C. Liu, H.M. Cheng, Phys. Rev. Lett., 85 (2000) 2617-2620.
[67] Z.M. Li, Z.K. Tang, H.J. Liu, N. Wang, C.T. Chan, R. Saito, S. Okada, G.D. Li, J.S. Chen, N. Nagasawa, S. Tsuda, Phys. Rev. Lett., 87 (2001) 127401.
[68] A. Hartschuh, H.N. Pedrosa, L. Novotny, T.D. Krauss, Science, 301 (2003) 1354-1356.
[69] M.F. Islam, D.E. Milkie, C.L. Kane, A.G. Yodh, J.M. Kikkawa, Phys. Rev. Lett., 93 (2004) 037404.
[70] L. Ren, C.L. Pint, L.G. Booshehri, W.D. Rice, X. Wang, D.J. Hilton, K. Takeya, I. Kawayama, M. Tonouchi, R.H. Hauge, J. Kono, Nano Lett., 9 (2009) 2610.
[71] C.L. Pint, Y.-Q. Xu, S. Moghazy, T. Cherukuri, N.T. Alvarez, E.H. Haroz, S. Mahzooni, S.K. Doorn, J. Kono, M. Pasquali, R.H. Hauge, ACS Nano, 4 (2010) 1131-1145.
[72] A. Gupta, G. Chen, P. Joshi, S. Tadigadapa, P.C. Eklund, Nano Lett., 6 (2006) 2667-2673.
[73] T. Mishina, K. Nitta, Y. Masumoto, Phys. Rev. B, 62 (2000) 2908-2911.
[74] K. Ishioka, M. Hase, M. Kitajima, L. Wirtz, A. Rubio, H. Petek, Phys. Rev. B, 77 (2008) 121402.
[75] J.-H. Kim, M.H. Jung, B.H. Hong, K.J. Yee, submitted, (2012).
[76] D.S. Lee, C. Riedl, B. Krauss, K.v. Klitzing, U. Starke, J.H. Smet, Nano Lett., 8 (2008) 4320-4325.
[77] A.C. Ferrari, J.C. Meyer, V. Scardaci, C. Casiraghi, M. Lazzeri, F. Mauri, S. Piscanec, D. Jiang, K.S. Novoselov, S. Roth, A.K. Geim, Phys. Rev. Lett., 97 (2006) 187401.
[78] G.G. Samsonidze, R. Saito, N. Kobayashi, A. Gruneis, J. Jiang, A. Jorio, S.G. Chou, G. Dresselhaus, M.S. Dresselhaus, Appl. Phys. Lett., 85 (2004) 5703.
[79] R.A. Jishi, L. Venkataraman, M.S. Dresselhaus, G. Dresselhaus, Chem. Phys. Lett., 77 (1993) 209.
[80] J. Jiang, R. Saito, A. Gruneis, S.G. Chou, G.G. Samsonidze, A. Jorio, G. Dresselhaus, M.S. Dresselhaus, Phys. Rev. B, 71 (2005) 205420.
[81] J. Jiang, R. Saito, A. Gruneis, G. Dresselhaus, M.S. Dresselhaus, Carbon, 42 (2004) 3169.
[82] R. Saito, K. Sato, Y. Oyama, J. Jiang, G.G. Samsonidze, G. Dresselhaus, M.S. Dresselhaus, Phys. Rev. B, 72 (2005) 153413.
[83] M. Machon, S. Reich, H. Telg, J. Maultzsch, P. Ordejon, C. Thomsen, Phys. Rev. B, 71 (2005) 035416.
[84] K. Sasaki, R. Saito, G. Dresselhaus, M.S. Dresselhaus, H. Farhat, J. Kong, Phys. Rev. B, 78 (2005) 235405.
[85] J. Jiang, R. Saito, G.G. Samsonidze, S.G. Chou, A. Jorio, G. Dresselhaus, M.S. Dresselhaus, Phys. Rev. B, 72 (2005) 235408.
[86] D. Porezag, T. Frauenheim, T. Kohler, G. Seifert, R. Kaschner, Phys. Rev. B, 51 (1995) 12947.
[87] T. Ando, H. Suzuura, Phys. Rev. B, 65 (2002) 235412.
[88] R. Saito, M.S. Dresselhaus, G. Dresselhaus, Phys. Rev. B, 61 (2000) 2981.
[89] K. Nakada, M. Fujita, G. Dresselhaus, M.S. Dresselhaus, Phys. Rev. B, 64 (1996) 17954.
[90] Y.W. Son, M.L. Cohen, S.G. Louie, Phys. Rev. Lett., 97 (2006) 216803.
[91] M. Fujita, K. Wakabayashi, K. Nakada, K. Kusakabe, J. Phys. Soc. Jpn, 65 (1996) 1920.